\documentclass[aps,prc,superscriptaddress,nofootinbib,floatfix,preprint,tightenlines]{revtex4-1}
 \usepackage{bbm}
  \usepackage{amsthm}
\usepackage{mathrsfs}
\usepackage{graphicx}
\usepackage{float} 
\usepackage{amsmath}
\usepackage{subfigure}
\usepackage{multirow}
\usepackage{amsfonts}
\newcommand{\eq}{\begin{eqnarray}}
\newcommand{\en}{\end{eqnarray}}
\newcommand{\Mhi}{M_{high}}
\newcommand{\Mlo}{M_{low}}

\begin{document}

\title{An alternative scheme for effective range corrections in pionless EFT}

\author{M.~Ebert}
\email{mebert@theorie.ikp.physik.tu-darmstadt.de}
\affiliation{Technische Universit\"at Darmstadt, Department of Physics,
64289 Darmstadt, Germany}

\author{H.-W.~Hammer}
\email{hans-werner.hammer@physik.tu-darmstadt.de}
\affiliation{Technische Universit\"at Darmstadt, Department of Physics,
64289 Darmstadt, Germany}
\affiliation{ExtreMe Matter Institute EMMI and Helmholtz Forschungsakademie
  Hessen f\"ur FAIR (HFHF),
GSI Helmholtzzentrum f\"ur Schwerionenforschung GmbH,
64291 Darmstadt, Germany}

\author{A.~Rusetsky}
\email{rusetsky@hiskp.uni-bonn.de}
\affiliation{HISKP and BCTP, Rheinische Friedrich-Wilhelms Universit\"at Bonn, 53115 Bonn, Germany}
\affiliation{Tbilisi State  University,  0186 Tbilisi, Georgia\vspace*{.3cm}}

\vspace*{.3cm}

\begin{abstract}
  We discuss an alternative scheme for including  effective range corrections
  in pionless effective field theory. The standard approach treats
  range terms as perturbative insertions in the $T$-matrix. In a finite
  volume this scheme can lead to singular behavior close to the
  unperturbed energies. We consider an alternative scheme that
  resums the effective range but expands the spurious pole of the
  $T$-matrix created by this resummation. We test this alternative
  expansion for several model potentials and observe good convergence.
\end{abstract}

\maketitle

\section{Introduction}

In nuclear physics, there is a hierarchy of Effective Field Theories (EFTs)
which all describe nuclear phenomena at
a certain resolution scale (for reviews see, e.g.,
Refs.~\cite{Bedaque:2002mn,Epelbaum:2008ga,Hammer:2019poc}.)
Pionless EFT describes the interactions of individual nucleons
at momenta small compared to the pion mass \cite{vanKolck:1997ut,Kaplan:1998tg,Kaplan:1998we,vanKolck:1998bw,Chen:1999tn}.
Apart from electroweak interactions, the effective Lagrangian
contains only short-range contact interactions between non-relativistic nucleons.
It can be understood as an expansion around the unitary limit of infinite
scattering length.  The  breakdown scale of pionless EFT
is set by the pion mass, $\Mhi\sim M_\pi$, while the typical 
low-energy scale is $\Mlo \sim 1/a \sim k$.
For momenta $k\sim M_\pi$, pion exchange can no longer be treated as
a short-range interaction and has to be included explicitly. This 
leads to chiral EFT whose breakdown scale $\Mhi$ is set by the chiral
symmetry breaking scale $\Lambda_\chi$~\cite{Weinberg:1990rz,Weinberg:1991um}.
The pionless theory exploits the large scattering length but 
is independent of the mechanism responsible for it. 
Thus it can be applied to a variety of systems ranging from
ultracold atoms to hadrons and nuclei.

At leading order (LO), one needs to resum a momentum-independent
contact interaction in order to describe the large scattering length
physics. This resummation is conveniently implemented using
dibaryon or dimer fields~\cite{Kaplan:1996nv}. At next-to-leading order (NLO)
the two-body ranges have to be included perturbatively.
In the dimer framework
this requires one insertion of the dimer kinetic-energy operator between LO
amplitudes.  At higher orders, the procedure of
perturbative range insertions becomes tedious, and a direct calculation of the
corrections requires fully off-shell LO amplitudes.  To avoid this, range
corrections can be resummed by including the effective range in the denominator
of the dimer propagator.   Early on it was noted that
this resummation introduces a spurious pole in the deuteron
propagator~\cite{Bedaque:1997qi}.
Located at a momentum scale of roughly
200~MeV, it is outside the range of validity of the EFT and thus in
principle is an irrelevant UV artifact.
However, in three- and higher-body systems it can limit the range of
cutoffs that can be
used in the numerical solution of the scattering equations.
In the three-nucleon system, this is especially
true in the doublet S-wave of neutron-deuteron scattering
(triton channel) unless measures are taken to remove the pole.
In the quartet S-wave, due to the Pauli principle, the solution is not
sensitive to this deep pole and the cutoff can be made arbitrarily large.

In Ref.~\cite{Bedaque:2002yg} it was proposed to partially re-expand
the resummed propagators and to use terms up to order $n$ for a calculation at
N$^n$LO.  Using these ``partially resummed'' propagators generates all desired
terms at a given order, but still retains some higher-order corrections, which
have to be small.\footnote{We note that it is important to keep the cutoff
at or below the breakdown scale of the theory to satisfy this requirement.}
The first strictly perturbative NLO calculation of $nd$ scattering in the
doublet $S$ channel was carried out in \cite{Hammer:2001gh}, implementing
the procedure suggested in \cite{Bedaque:1998km}. Ji et
al.~\cite{Ji:2011qg,Ji:2012nj} extended these calculations to N$^2$LO
and pointed out that an additional three-body term enters at NLO when
the scattering length is varied. This is particularly relevant
for applications in ultracold atoms and quark mass extrapolations. Finally,
Vanasse \cite{Vanasse:2013sda} developed a scheme that avoids the numerically
expensive determination of full off-shell amplitudes made in previous
perturbative calculations. Overall, he obtains
$nd$ phase shifts at N$^2$LO which are in good agreement with the empirical
behavior up to laboratory energies of $\simeq$~ 24~MeV.

In this paper, we revisit the problem of range corrections in the
three-body system from the perspective of the three-body quantization
condition in a finite volume, following the formalism developed in 
Refs.~\cite{Hammer:2017uqm,Hammer:2017kms}, see
also~\cite{Hansen:2014eka,Hansen:2015zga,Mai:2017bge} for alternative formulations.
For simplicity, we focus on the three-boson
system, which is known to have the same qualitative features as the 
neutron-deuteron doublet $S$-wave channel.
We expect the approach of \cite{Vanasse:2013sda} to be problematic
numerically in a finite volume. Indeed, in a finite box of size $L$,
the S-wave dimer propagator gets replaced by~\cite{Hammer:2017uqm,Hammer:2017kms}:
\eq\label{eq:tauL}
\tau_L({\bf k},{k^*}^2)=\frac{1}{k^*\cot\delta(k^*)+S({\bf k},{k^*}^2)}\, .
\en
Here, ${\bf k},k^*$ denote the total three-momentum of a dimer and
the magnitude of the relative momentum of two particles, constituting
a dimer, in their center-of-mass frame. Furthermore, $\delta(k^*)$ denotes
the pertinent phase shift and the quantity $S({\bf k},{k^*}^2)$ stands for
the infinite sum
\eq\label{eq:S}
S({\bf k},{k^*}^2)=-\frac{4\pi}{L^3}\,\sum_{\bf p}\frac{1}{{\bf p}^2+{\bf p}{\bf k}+{\bf k}^2-mE}\, ,\quad\quad {\bf p}=\frac{2\pi}{L}\,{\bf n}\, ,\quad
{\bf n}=\mathbb{Z}^3\, ,
\en
where $E$ is the total energy of the particle-dimer system in the rest frame.\footnote{This sum diverges and has to be properly regularized, e.g., by using dimensional regularization. The details can be found in Refs.~\cite{Hammer:2017uqm,Hammer:2017kms}.}
In the infinite volume, the sum turns into the integral that can be easily evaluated, leading to a well-known result.

The problem with expanding the finite-volume dimer propagator in a manner
proposed in
Refs.~\cite{Hammer:2001gh,Bedaque:1998km,Ji:2011qg,Ji:2012nj,Vanasse:2013sda}
is related to the singularities of the denominator. Namely,
from Eqs.~(\ref{eq:tauL}) and (\ref{eq:S})
it can be immediately seen that, in a finite volume,
the propagator has an infinite tower of poles above the elastic threshold,
corresponding to the finite-volume energy spectrum in the two-particle
subsystems. In the infinite volume, these poles condense and form an elastic
cut. Next, we note that, in a finite volume,
the expansion will not work in the vicinity of these
poles, producing denominators that are more and more singular.
Bearing this fact in mind, we aim at an alternative procedure for removing
the spurious poles, which
is not based on such an expansion and, hence,
high powers of the energy denominator never appear. Below, we shall
demonstrate, how this goal can be achieved.

The paper is organized as follows. In In Sect.~\ref{sec:notations}
we set up the EFT framework which allows one to study the three-particle
problem in a systematic manner. In Sect.~\ref{sec:formalism} we formulate
a method that allows one to consistently remove a spurious subthreshold
pole from the dimer propagator. In Sect.~\ref{sec:numerics} this method
is numerically tested within a toy model. The convergence of the approach,
as well as the applicability of the power counting is discussed in detail.
Finally, Sect.~\ref{sec:concl} contains our conclusions.

\section{Formalism}
\label{sec:notations}
\subsection{Non-relativistic Lagrangians}

We consider the system of three identical non-relativistic bosons with a mass $m$, described
by the field $\psi$.
In this system a non-derivative three-body interaction is required for
renormalization already at leading order~\cite{Bedaque:1998kg}.
The Lagrangian takes the form (only S-wave contributions are shown explicitly):
\eq\label{eq:particle}
\mathscr{L}&=&\psi^\dagger\biggl(i\partial_0+\frac{\nabla^2}{2m}\biggr)\psi
-\frac{C_0}{2}\,(\psi^\dagger\psi)^2
+\frac{C_2}{4}\,\biggl((\psi^\dagger{\stackrel{\leftrightarrow}{\nabla}}^2\psi^\dagger)\psi^2+\mbox{h.c.}\biggr)
\nonumber\\[2mm]
&-&\frac{D_0}{6}\,(\psi^\dagger\psi)^3
-\frac{D_2}{9}\,
\biggl((\psi^\dagger{\stackrel{\leftrightarrow}{\nabla}}^2\psi^\dagger)\psi^\dagger\psi^3+\mbox{h.c.}\biggr)+\cdots\, ,
\en
where $\stackrel{\leftrightarrow}{\nabla}=\frac{1}{2}\,
(\stackrel{\rightarrow}{\nabla}-\stackrel{\leftarrow}{\nabla})$
is a Galilei-invariant derivative. The couplings $C_0,\,C_2$,
describe the interactions in the two-particle sector and can be related to
the S-wave
scattering length $a$ and effective range $r$, respectively.
$D_0$ and $D_2$ correspond to 
three-body interactions with zero/two derivatives.
Higher-order terms with more derivatives
are not shown explicitly.

To describe the three-body systems, it is convenient to work in the
particle-dimer formalism.  The dimer
can be introduced as an auxiliary integration variable in the path integral.
In this manner, it is obvious that the theory with dimers leads to the same
Green functions.
The particle-dimer Lagrangian takes the form\footnote{See, e.g., Refs.~\cite{Kaplan:1996nv,Bedaque:1998km,Bedaque:2002yg}.}
\eq\label{eq:dimer}
\mathscr{L}_d&=&\psi^\dagger\biggl(i\partial_0+\frac{\nabla^2}{2m}\biggr)\psi
+\sigma d^\dagger\biggl(i\partial_0+\frac{\nabla^2}{4m}+\Delta\biggr)d
+\frac{f_0}{2}\,(d^\dagger\psi^2+\mbox{h.c.})+\cdots
\nonumber\\[2mm]
&+&h_0d^\dagger d\psi^\dagger\psi +h_2d^\dagger d(\psi^\dagger\nabla^2\psi+(\nabla^2\psi^\dagger)\psi)+\cdots\, .
\en
Here, the ellipses stand for the terms that contain more space derivatives or higher partial waves, $d$ denotes the dimer field, and the sign $\sigma=\pm 1$ determines the sign of the effective range. In the examples discussed below,
we have $\sigma=-1$. The two
Lagrangians (\ref{eq:particle}) and (\ref{eq:dimer}) describe the same physics, so the couplings can be matched to each other. This matching has been
considered in the literature many times (see, e.g., Refs.~\cite{Bedaque:1999vb,Braaten:2004rn}) and we do not repeat it here once more.
Note only that two couplings $C_0,C_2$ (or, the scattering length and the effective range) can be traded for
two parameters $\Delta,f_0$, whereas two other couplings $D_0,D_2$ can be expressed
through $h_0,h_2$.

In the dimer picture, the three-particle amplitude is expressed through the particle-dimer
amplitude in a closed form. The latter obeys an integral equation (the Faddeev or Skorniakov-Ter-Martirosian equation), which can be readily obtained, considering the diagrammatic expansion of the amplitude. Note that the dimer need not correspond to a physical particle. Within this approach,
it is just a useful mathematical tool that makes the bookkeeping of various diagrams
extremely simple. In the numerical study that follows, however, we shall adjust the parameters so that the dimer is a stable particle, and use parameter values from the two-nucleon systems. The on-shell particle-dimer scattering
amplitude then has a direct physical interpretation.

\subsection{Faddeev equation for the particle-dimer scattering}

As already mentioned, the particle-dimer scattering amplitude in the non-relativistic effective theory obeys
the Faddeev equation
\eq\label{eq:BS}
M({\bf p},{\bf q};E)=Z({\bf p},{\bf q};E)
+8\pi\int^\Lambda\frac{d^3{\bf k}}{(2\pi)^3}\,Z({\bf p},{\bf k};E)\tau({\bf k};E)M({\bf k},{\bf q};E)\, ,
\en
where $E$ is the total energy of the particle-dimer system in the center-of-mass (CM) frame, and $\tau({\bf k};E)$ denotes the two-body amplitude.
It is always assumed that $E$ has an infinitesimal positive imaginary part
$E\to E+i \varepsilon$. As in the Lagrangian (\ref{eq:dimer}), we have
included only S-wave two-body interactions. Higher-partial wave
interactions contribute only beyond the order considered here.
The S-wave two-body amplitude in Eq.~(\ref{eq:BS}) is given by:
\eq
\tau({\bf k};E)\doteq \tau(k^*)=\biggl(k^*\cot\delta(k^*)+k^*\biggr)^{-1}\, ,
\en
where $\delta(k^*)$ denotes the S-wave phase shift, and $k^*$ is the magnitude of the
boosted relative momentum. In the non-relativistic kinematics,
\eq
k^*=\sqrt{\frac{3}{4}\,{\bf k}^2-mE}\, .
\en
Here, $m$ stands for the particle mass. Further, {\em for small momenta,} the effective-range
expansion can be carried out:
\eq
k^*\cot\delta(k^*)=-\frac{1}{a}-\frac{1}{2}\,r{k^*}^2+\cdots\, ,
\label{Propagator}
\en
where $a$ and $r$ stand for the scattering length and the effective range, respectively.

The kernel in the Faddeev equation consists of the one-particle
exchange contribution and a tower of the polynomial terms with the increasing powers of momenta, which are obtained from the
particle-dimer interaction Lagrangian:
\eq\label{eq:Z}
Z({\bf p},{\bf q};E)=\frac{1}{{\bf p}^2+{\bf q}^2+{\bf p}{\bf q}-mE}
+\frac{H_0}{\Lambda^2}+\frac{3H_2}{8\Lambda^4}\,({\bf p}^2+{\bf q}^2)\, ,
\en
where the parameters $H_0,H_2,\ldots$ can be expressed in terms of the
effective couplings in the Lagrangian $h_0,h_2,\ldots$. Further, $H_0,H_2,\ldots$ depend on the cutoff
$\Lambda$ so that the scattering amplitude $M({\bf p},{\bf q};E)$ is $\Lambda$-independent at a given order in the low-energy expansion.

Carrying out a partial-wave expansion in the Faddeev equation
and projecting onto S-wave results into:
\eq
M(p,q;E)=Z(p,q;E)
+\frac{4}{\pi}\,\int^\Lambda k^2dk\,
Z(p,k;E)\tau(k^*)M(k,q;E)\,,
\label{FaddeevSWave}
\en
where
\eq
Z(p,q,E)=\frac{1}{2pq}\,\ln\frac{p^2+q^2+pq-mE}{p^2+q^2-pq-mE}
+\frac{H_0}{\Lambda^2}+\frac{3H_2}{8\Lambda^4}\,( p^2+q^2)+\cdots\, ,
\en
where the subscript $\ell=0$ has been
dropped in all amplitudes. Note that this has been done only
in order to keep the formulae
simple and transparent. If needed, the formalism can be easily extended
to include higher partial waves (see, e.g., Ref.~\cite{Hammer:2017kms}).

Further,
as shown in Ref.~\cite{Bedaque:2002yg}, introducing a trimer auxiliary
field in the
Lagrangian along with the dimer field,
is it possible to simplify the Faddeev equation. In the kernel of the
transformed equation, the three-momenta are traded
for the total energy $E$:
\eq\label{eq:ZE}
Z(p,q,E)\to\frac{1}{2pq}\,\ln\frac{p^2+q^2+pq-mE}{p^2+q^2-pq-mE}
+\frac{H_0}{\Lambda^2}+\frac{H_2}{\Lambda^4}\,(mE+\gamma^2)+\cdots\, ,
\en
where $\gamma=\sqrt{mE_d}$ and $E_d$ denotes the binding energy of the dimer.
The amplitude, which is a solution of the equation with the transformed
kernel, is equal to the original amplitude up to the higher-order terms.
It is slightly easier to use the transformed kernel in numerical calculations
and we shall stick to this option in the following.

In the presence of a stable dimer, the on-shell amplitude $M$ is related to the
particle-dimer scattering phase, according to:
\eq\label{eq:EFTphase0}
M(p,p,E_p)=\frac{3}{16\gamma}\,\frac{1}{p\cot\delta(p)-ip}\, ,
\en
This phase is real below the dimer breakup threshold at $E=0$.

\section{Problem and Solution}
\label{sec:formalism}

\subsection{Spurious states}

The hard scale $M_{high}$ in the two-body interactions is set by the effective range $r$.
To make further discussion as transparent as possible, let us assume that
true dynamics of a system, which {\em at small momenta} is described
by the non-relativistic effective Lagrangian, is such that no deeply bound two-body
states with $\sqrt{mE_2}\simeq |r|^{-1}$ emerge. The effective field theory
setting in the present form could not be used to consistently describe such states anyway,
and we merely discard them (in the two-particle sector, the presence of such states at small momenta
will show up only indirectly, through their contributions to the effective couplings).
Only shallow bound states with $\sqrt{mE_2}\ll |r|^{-1}$ will be allowed. In particular,
in the following, we shall tune our parameters so that only one shallow bound state
-- a dimer -- with the binding energy $E_d>0$ exists. Hence, the two-body scattering
length $a$ must be large and positive, $a\gg |r|$.

After this introduction, let us formulate the problem. If in the Faddeev
equation~(\ref{eq:BS}), the integration momentum $|{\bf k}|$ runs from $0$ to
$\Lambda$, the quantity $k^*$ varies from $k^*=\sqrt{-mE}$ to
$k^*\simeq \frac{\sqrt{3}}{2}\,\Lambda$ (if $E<0$, the quantity $k^*$ is always real).
Thus, the subthreshold amplitude at large momenta enters
the equation. In the effective theory, all that can be done is to approximate
$k^*\cot\delta(k^*)$ by means of the effective range expansion, which
does not make sense at large momenta. One would argue that the behavior at large
momenta should not really matter and can be taken care of by an appropriate
renormalization prescription. Hence, it would be harmless to extend the integration
to high momenta. In reality, however, the situation is more subtle.
Let us retain only the first two terms in the effective-range expansion. Then, if $r>0$,
the two-body amplitude $\tau(k^*)$ develops a {\em spurious pole} at large momenta:
\eq\label{eq:twopole}
\tau(k^*)=\frac{1}{-1/a-r{k^*}^2/2+k^*}=\frac{-2/r}{(k^*-k_1)(k^*-k_2)}\, ,
\en
where
\eq
k_1=\frac{2/a}{1+\sqrt{1-2r/a}}\simeq \frac{1}{a}\, ,\quad\quad
k_2=\frac{1+\sqrt{1-2r/a}}{r}\simeq \frac{2}{r}\, .
\en
It is obvious that $k_1$ and $k_2$ correspond to the physical dimer
and to a spurious deep pole, respectively. Such a spurious pole emerges, because effective range expansion
is applied in a region where it is not valid anymore. Including higher orders in the expansion
will generate even more spurious poles.

An immediate consequence of the emerging spurious pole is that the integration
contour hits a singularity where, originally, there was no singularity. It should be
understood that the presence of the singularity is not a problem {\it per se:} in fact, in a theory where
physical deeply bound states are present, there are also singularities and one has
to handle them by deforming the integration contour or otherwise.
On the other hand, the fact that such a spurious pole contributes to the
unitarity relation
is a true problem. If in reality there is no such state, there should be no
such contribution at all. Even worse, in the case relevant to the two-nucleon
problem, $a>0$ and $r>0$,
the spurious pole has a residue with a wrong sign,
leading to the negative probabilities. Indeed, as can be
seen from Eq.~(\ref{eq:twopole}), the residues at two poles have opposite signs and,
since the dimer corresponds to a true bound state, the second pole has to correspond
to a ghost. In addition, it is not immediately
clear, how such contributions can be removed by changing the renormalization prescription
for the effective couplings, which are presumed to be real.

In the literature, one encounters different prescriptions for treating such spurious
singularities. For example, one may keep the cutoff $\Lambda$ low enough, so that the
spurious poles do not appear on the integration path.
The shortcomings of this approach, both conceptual and practical, are obvious.
First of all, one cannot remove the cutoff and ensure the independence of the results
on the regularization. Moreover, the upper bound of the cutoff depends on the order
one is working, and on the values of the effective-range expansion parameters.
Hence, setting up a universal upper bound is not possible in general.

The power counting of pionless EFT stipulates that the effective range
corrections in the three-body system are perturbative, since $|a|\gg|r|$
\cite{Bedaque:1998km}. This approach is implemented in 
Refs.~\cite{Ji:2011qg,Vanasse:2013sda}. It is reminiscent of the threshold expansion
of Beneke and Smirnov~\cite{Beneke:1997zp} (see also Refs.~\cite{Mohr:2003du,Mohr:2005pv}), and the heavy baryon expansion
in Chiral Perturbation Theory~\cite{Jenkins:1990jv,Mannel:1991mc,Bernard:1992qa}. This approach is based on the observation that
the Taylor-expansion of the propagators alters only high-momentum contributions in
the Feynman graphs -- exactly those, which are responsible for the trouble. Namely,
following Refs.~\cite{Ji:2011qg,Vanasse:2013sda}, one may expand the quantity $\tau(k^*)$, given by
Eq.~(\ref{eq:twopole}), in series in the effective range $r$ and include the
contributions in strict perturbation theory. The energy denominators,
$(-1/a+k^*)^{-n}$, obtained in a result of this expansion, do not produce spurious poles.
The resulting Faddeev equation can be readily solved -- the solution is written
down as a series in powers of the effective range parameter $r$. The method is very
appealing, successful, and fully consistent. However, using this method in a finite volume, following
the approach of Refs.~\cite{Hammer:2017uqm,Hammer:2017kms}, is not very convenient numerically, since the
denominator in a finite volume becomes very singular (cf. the discussion
in the introduction). For this reason,
in this paper we propose an alternative approach to this problem, where only the spurious
pole contribution is expanded. In this manner, high powers of energy denominator never appear. In addition, in our opinion, this method could be even simpler in the applications.

\subsection{Method}

Let us in the beginning assume that we work below the dimer breakup threshold,
$E<0$. The argument is then crystal clear.
We start by splitting off two poles in Eq.~(\ref{eq:twopole}) from each other:
\eq
\tau(k^*)=\frac{2(k_1+k_2)/r}{(k_2-k_1)(k^*+k_2)(k^*-k_1)}
-\frac{4k_2/r}{(k_2-k_1)({k^*}^2-k_2^2)}\, .
\en
Here, the first/second term contain the dimer/spurious poles, respectively. Note now
that the second term is, in fact, a low-energy polynomial -- since $k_2$ is a quantity of order of a heavy scale, $k_2\sim M_{high}$, 
it can be expanded in Taylor series in ${k^*}^2$. Doing this, one gets
rid of the spurious pole. It should be however demonstrated that the change in the
amplitude, which results by replacing the deep pole by its Taylor expansion,
can be indeed accounted for by adjusting the effective couplings. Below,
we shall demonstrate this by explicit calculations at one loop and
interpret this adjustment physically.

It is convenient to introduce the notations:
\eq
f(k^*)=-\frac{4k_2/r}{(k_2-k_1)({k^*}^2-k_2^2)}
-\frac{4k_2/r}{(k_2-k_1)k_2^2}\biggl\{1+\frac{{k^*}^2}{k_2^2}+\frac{{k^*}^4}{k_2^4}+\cdots\biggr\}\, ,
\label{f}
\en
as well as
\eq
f_1(k^*)&=&-\frac{4k_2/r}{(k_2-k_1)({k^*}^2-k_2^2)}
-\frac{4k_2/r}{(k_2-k_1)k_2^2}\, ,
\nonumber\\[2mm]
f_2(k^*)&=&-\frac{4k_2/r}{(k_2-k_1)({k^*}^2-k_2^2)}
-\frac{4k_2/r}{(k_2-k_1)k_2^2}\biggl\{1+\frac{{k^*}^2}{k_2^2}\biggr\}\, ,
\nonumber\\[2mm]
f_3(k^*)&=&-\frac{4k_2/r}{(k_2-k_1)({k^*}^2-k_2^2)}
-\frac{4k_2/r}{(k_2-k_1)k_2^2}\biggl\{1+\frac{{k^*}^2}{k_2^2}+\frac{{k^*}^4}{k_2^4}\biggr\}\, ,
\label{f123}
\en
and so on.
In other words, from the term corresponding to the spurious pole, we subtract its Taylor
expansion, up to some order. Further, writing down
$\tau(k^*)=[\tau(k^*)-f(k^*)]+f(k^*)$,
the Faddeev equation can be rewritten in the following form:
\eq
M({\bf p},{\bf q};E)&=&W({\bf p},{\bf q};E)
+8\pi\int^\Lambda\frac{d^3{\bf k}}{(2\pi)^3}\,W({\bf p},{\bf k};E)
[\tau(k^*)-f(k^*)]M({\bf k},{\bf q};E)\, ,
\nonumber\\[2mm]
W({\bf p},{\bf q};E)&=&Z({\bf p},{\bf q};E)
+8\pi\int^\Lambda\frac{d^3{\bf k}}{(2\pi)^3}\,Z({\bf p},{\bf k};E)f(k^*)W({\bf k},{\bf q};E)\, .
\label{ChangedFaddeev}
\en
Note now that in the first equation of the above system, which determines the
amplitude $M$ one is looking for, the spurious pole is replaced by its Taylor expansion.
Consequently, the culprit has been removed. The question remains, however, whether
the effective potential $W$, which is determined by the second equation, has the same
properties as $Z$, i.e., is given by a sum of the one-particle exchange diagram
and a low-energy polynomial. In this case, one could forget about the second equation
at all, since the difference between $W$ and $Z$ could be accounted for by a change of
the renormalization prescription.

In the following, we expand the quantity $W$ in the Born series
\eq\label{eq:PT}
W=Z+ZfZ+ZfZfZ+\cdots =W^{(1)}+W^{(2)}+W^{(3)}+\cdots\, ,
\en
in order to study the structure of each term separately. In particular, considering a couple
of simple examples at the second order, we verify that $W^{(2)}$ has indeed the
structure which was conjectured from the beginning. The generalization to higher
orders is clear.

Let us now start with the calculation of $W^{(2)}$. The quantity $Z$, displayed in
Eq.~(\ref{eq:Z}), contains an infinite number of terms, and hence $W^{(2)}$ will contain
infinite number of all cross products. To illustrate our statement, we pick out
a single term. The simplest choice is the one, proportional to $H_0^2$:
\eq
W^{(2)}_{00}&=&-\frac{32\pi k_2/r}{k_2-k_1}\,\biggl(\frac{H_0}{\Lambda^2}\biggr)^2
I_{00}\, ,
\nonumber\\[2mm]
I_{00}&=&
\int^\Lambda \frac{d^3{\bf k}}{(2\pi)^3}\,
\biggl\{\frac{1}{{k^*}^2-k_2^2-i\varepsilon}+\frac{1}{k_2^2}\biggl(
1+\frac{{k^*}^2}{k_2^2}+\cdots\biggr)\biggr\}\, .
\en
Note that the sign of $i\varepsilon$ follows from the prescription $E\to E+i\varepsilon$.

The imaginary part of $I_{00}$ is a constant, which depends on the energy $E$:
\eq
\mbox{Im}\,I_{00}=\frac{2}{3\sqrt{3}\pi}\,\sqrt{k_2^2+mE}=
\frac{2k_2}{3\sqrt{3}\pi}\,\biggl\{1+\frac{mE}{k_2^2}+\cdots\biggr\}\, .
\en
We assume here that the cutoff $\Lambda$ is chosen large enough, so the pole
is inside the integration region -- otherwise, the imaginary part would vanish.
Further, the real part is also a low-energy polynomial:
\eq
\mbox{Re}\,I_{00}=\frac{2}{3\pi^2}\,\Lambda+\frac{1}{2\pi^2k_2^2}\,\biggl(
\frac{1}{3}\,\Lambda^3+\frac{1}{k_2^2}\biggl(\frac{3}{20}\Lambda^5
-\frac{1}{3}\,\Lambda^3mE\biggr)+\cdots\biggr\}\, .
\en
It can be seen that the real part can be removed by altering the renormalization
prescription. The sole subtle point is that the counterterms depend on (are low-energy
polynomials of) the total three-particle CM energy $E$ which, in the Lagrangian,
translates into time derivatives on both the particle and dimer fields.
The following discussion
demonstrates, how one could circumvent this problem. First, if one is interested only
in the on-shell particle-dimer scattering matrix, one could directly use the equations of
motion (EOM) in the particle-dimer Lagrangian, trading the time derivatives for space
ones. In the description of the generic three particle processes, however, the dimers
may go off-shell. In this case, one should first integrate the dimer field out and then
use the EOM for the particle fields that leaves the three-body $S$-matrix elements
unchanged.

Applying the same procedure to the imaginary part leads, however, to a conceptual
inconsistency, since the counterterms, which are needed to remove it, should be complex.
The problem with the spurious poles shows up exactly at this place. Note, for example,
that if the cutoff $\Lambda$ is chosen so small that the integration contour does not
hit the pole, then the problem does not arise, since the imaginary part vanishes.
It is also clear that one could circumvent the problem, which originates from the use of
the effective-range expansion beyond the range of its applicability, by merely
dropping the imaginary part by hand (because, in the exact theory, there are no poles
and thus no imaginary part).

As a side remark, this discussion also shows how physical deep bound
states should be treated.
The corresponding poles are physical and cannot be eliminated from the
theory. On the other hand, it would be inconsistent to treat them in the present setting
explicitly, because their binding energy is determined by the
hard scale $M_{high}$. According to the above
discussion, such a deep bound state pole will show up indirectly, through the contribution
to the effective couplings, which become complex. In contrast to the case of spurious
poles, the imaginary part corresponds to the contribution of the physical
deep bound state to the unitarity relation and cannot be discarded. The potential $W$ becomes now a
kind of ``optical potential''~\cite{optical}, in which the shielded states manifest
themselves through the imaginary part. It should be also mentioned
that the contribution from the physical states to the imaginary part always comes with a
correct sign, in accordance with unitarity.

Next, we shall consider another contribution to the quantity $W^{(2)}$ that will allow us
to have a closer look on its structure at small momenta. We shall namely single
out the term where both $Z$ are replaced by the one-particle exchange
contribution:
\eq
W^{(2)}_{ee}&=&-\frac{32\pi k_2/r}{k_2-k_1}\,(I_{\sf pole}-I_{\sf subtr})\, ,
\nonumber\\[2mm]
I_{\sf pole}&=&\frac{4}{3}\,\int^\Lambda\frac{d^3{\bf k}}{(2\pi)^3}\,
  \frac{1}{{\bf p}^2+{\bf p}{\bf k}+{\bf k}^2-mE-i\varepsilon}\,
  \frac{1}{{\bf k}^2-\rho^2-i\varepsilon}\,
\nonumber\\[2mm]
  &\times& \frac{1}{{\bf k}^2+{\bf k}{\bf q}+{\bf q}^2-mE-i\varepsilon}\, ,
\nonumber\\[2mm]
  I_{\sf subtr}&=&-\frac{1}{k_2^2}\,\int^\Lambda\frac{d^3{\bf k}}{(2\pi)^3}\,
  \frac{1}{{\bf p}^2+{\bf p}{\bf k}+{\bf k}^2-mE-i\varepsilon}\,
  \biggl(1+\frac{{k^*}^2}{k_2^2}+\cdots\biggr)
  \nonumber\\[2mm]
  &\times&\frac{1}{{\bf k}^2+{\bf k}{\bf q}+{\bf q}^2-mE-i\varepsilon}\, ,
  \en
  and $\rho^2=\frac{4}{3}\,(k_2^2+mE)$.

  The integral $I_{\sf pole}$ is ultraviolet-finite, and hence the cutoff $\Lambda$ can
  be taken to infinity. Using the Feynman trick, it can be written in the following form:
  \eq
  I_{\sf pole}=\frac{1}{12\pi}\,\int_0^1dx \int_0^1ydy
  \frac{1}{(A+By+Cy^2-i\varepsilon)^{3/2}}\, ,
  \en
  where
  \eq
  A&=&-\rho^2\, ,
  \nonumber\\[2mm]
  B&=&-mE+\rho^2+x{\bf p}^2+(1-x){\bf q}^2\, ,
  \nonumber\\[2mm]
  C&=&-\frac{1}{4}\,(x{\bf p}+(1-x){\bf q})^2\, .
  \en
  The integral over the variable $y$ can be performed, yielding:
  \eq\label{eq:pole}
  I_{\sf pole}&=&\frac{1}{6\pi}\,\int_0^1dx\frac{1}{4AC-B^2}\,\biggl(
\frac{2A}{(-\rho^2-i\varepsilon)^{1/2}}
\nonumber\\[2mm]
  &-&\frac{2A+B}{(-mE+x{\bf p}^2+(1-x){\bf q}^2-\frac{1}{4}\,(x{\bf p}+(1-x){\bf q})^2-i\varepsilon)^{1/2}}\biggr)\, .
    \en
    The first term is again a low-energy polynomial (with complex coefficients) and can
    therefore be discarded, while the second term is not. Expanding the numerator in the
    integrand in a Taylor series, we get: 
    \eq\label{eq:ABC}
    -\frac{2A+B}{4AC-B^2}&=&-\frac{3}{4k_2^2}
\nonumber\\[2mm]
    &-&\frac{3}{16k_2^4}\,\biggl(5mE-9(x{\bf p}^2+(1-x){\bf q}^2)+3(x{\bf p}+(1-x){\bf q})^2\biggr)+\cdots\, .
    \en
    Next, consider the subtraction integral
    $I_{\sf subtr}=\sum_n I_{\sf subtr}^{(n)}/k_2^{2n}$.
    The leading term is given by:
    \eq
  I_{\sf subtr}^{(1)}&=&  
\int^\Lambda\frac{d^3{\bf k}}{(2\pi)^3}\,
  \frac{1}{{\bf p}^2+{\bf p}{\bf k}+{\bf k}^2-mE-i\varepsilon}\,
  \frac{1}{{\bf k}^2+{\bf k}{\bf q}+{\bf q}^2-mE-i\varepsilon}
\nonumber\\[2mm]
&=&\frac{1}{8\pi}\,\int_0^1dx\frac{1}
{(-mE+x{\bf p}^2+(1-x){\bf q}^2-\frac{1}{4}\,(x{\bf p}+(1-x){\bf q})^2-i\varepsilon)^{1/2}}\, .
  \en
  It is immediately seen that the leading-order term $I_{\sf subtr}^{(1)}$ cancels the leading-order non-polynomial piece in  $I_{\sf pole}$ that emerges from the first term
  in the expansion in Eq.~(\ref{eq:ABC}). The higher-order terms such as
  \eq
  I_{\sf subtr}^{(2)}=-\frac{3\Lambda}{8\pi^2}
-\frac{1}{32\pi}\,\int_0^1dx\frac{5mE-9(x{\bf p}^2+(1-x){\bf q}^2)+3(x{\bf p}+(1-x){\bf q})^2}
{(-mE+x{\bf p}^2+(1-x){\bf q}^2-\frac{1}{4}\,(x{\bf p}+(1-x){\bf q})^2-i\varepsilon)^{1/2}}\, ,
\en
have the same property.
  The integral cancels against the next-to-leading order non-polynomial contribution, emerging from the second term in Eq.~(\ref{eq:ABC}), and only the polynomial contribution
  is left at this order. The role of the higher-order subtraction terms is similar --
  they merely remove the non-polynomial contributions at the pertinent order, leaving
  only the polynomial parts (as it should indeed be).
  
  The general pattern becomes crystal clear already from these examples,
  and there is no need to
  consider higher-order terms. To summarize,
  the quantity $W$ is indeed a low-energy polynomial
  up to an order fixed by the order of the subtracted polynomial. The coefficients of this
  polynomial are energy-dependent and complex. The energy-dependence can be
  eliminated through the use of the EOM. The imaginary parts, arising from the spurious
  poles, are artifacts of the use of the effective-range expansion for large momenta. Our
  prescription consists of dropping these artifacts since, in the full theory,
  there are no poles leading to the complex potential. Thus, one may finally
  assume that $W=Z$, modulo the change in the renormalization prescription.

  Final remarks about unitarity are in order. The un-expanded two-body
  amplitude, which still contains the spurious pole, obeys exact two-body
  unitarity by construction, whereas this property is lost after expansion.
  However, the violation is small in the physically relevant region of small
  momenta, because ${{k^*}^2}\!/k_2^2\sim M_{low}^2/M_{high}^2$ is a small parameter there.  Moreover, the violation of unitarity in this region
  can be systematically reduced,
  including higher-order terms in the Taylor expansion. 
  Further, our argument can be extended for the energies above the breakup
  threshold, $E>0$. In this region, it is no longer true that the contributions
  to the imaginary part of $W$ come solely from the spurious subthreshold
  pole. In fact, they can emerge also from the denominators, corresponding
  to the particle exchange between the dimer and spectator particle. This
  contribution to the imaginary part is physical and should be retained.
  Note, however, that this contribution emerges exclusively from the region
  of small integration momenta, where the quantity $k^*$ is small. In this
  region, the quantity $f(k^*)$ is also small (it converges to zero in the
  Taylor expansion in ${{k^*}^2}\!/k_2^2$).
  Hence, the corresponding contribution
  to the imaginary part of $W$ should be small. It can be systematically
  reduced by including higher-order terms in the Taylor expansion.
  Thus it can be safely neglected.

  It should also be mentioned that the relation of the amplitude to the phase shift is
  modified along with the unitarity relation, if the subtraction is done. In particular,
  instead of Eq.~(\ref{eq:EFTphase0}), one now has:
\eq\label{eq:EFTphase}
M(p,p,E_p)=\frac{3}{16\gamma}\,\frac{k_2-k_1}{k_2+k_1}\,\frac{1}{p\cot\delta(p)-ip}\, .
\en
Note that Eq.~(\ref{eq:EFTphase}) reduces to Eq.~(\ref{eq:EFTphase0}) in the limit
$r\to 0$, as it should.

  \subsection{Order of the subtraction polynomial}

  It is natural to ask how large the order of the subtracted polynomial
  in $f(k^*)$ should be. Is it so that, if one subtracts more terms, the accuracy of the method increases?  The answer to this question is obviously no. Recall that one has to compensate the subtraction by adjusting effective couplings in the Lagrangian. If one does not have enough couplings $H_0,H_2,\ldots$, a  further subtraction does not lead to an improved accuracy.

  Since the problem is highly non-perturbative, it is difficult to establish
  the order of the subtraction polynomial a priori without a non-perturbative
  calculation. We stress that the
  requirement to promote the three-body interaction to leading order
  in Ref.~\cite{Bedaque:1998kg} was also established by explicitly
  investigating the cutoff dependence of
  numerical solutions of Eq.~(\ref{FaddeevSWave}).
  Alternatively, one can analyze the asymptotic behavior of
  non-perturbative solutions \cite{Danilov:61,Griesshammer:2005ga}.
  In order to get a first idea on optimal number of subtractions, we 
  start with a perturbative analysis of Eq.~(\ref{FaddeevSWave}), being well
  aware of the shortfalls of this approach.

  It is convenient to consider
  the effective potential $W$, rather than the amplitude $M$. It is
  straightforward to establish counting rules for $W$ in perturbation theory. Indeed, assume that one is using dimensional
  regularization to tame ultraviolet divergences in this quantity (the use of
  any other regularization, say, the cutoff regularization, will alter only
  the polynomial part of $W$ that can be compensated by a choice of the renormalization prescription). Further, the quantity $Z$
  (containing the exchange diagram) counts
  at $O(p^{-2})$ for small three-momenta (all low-energy constants count at
  order $p^0$). The quantities $f_1(k^*),f_2(k^*),\ldots$, introduced in
  Eq.~(\ref{f123}), for small values of $k^*$ count as $O(p^2),O(p^4),\ldots$.
  Finally, the integration measure $d^3{\bf k}$ counts at $O(p^3)$.

  Let us now consider the perturbative expansion of the potential $W$, given
  by Eq.~(\ref{eq:PT}). Each consecutive term in this expansion contains one
  additional factor $Z$, $f(k^*)$ and  $d^3{\bf k}$ -- hence, the power in $p$
  increases at least by one, when one goes to higher-order terms.
  Hence, the most stringent constraint on the number of subtractions arises
  from the term $W^{(2)}$. At lowest order, one has to replace
  $f(k^*)$ by $f_1(k^*)$.
  Then, $W^{(2)}$ counts  at $O(p^{3-2+2-2})=O(p)$  according to our power
  counting. Of course, this counting concerns the non-analytic piece
  of $W^{(2)}$ only. Furthermore, taking $f_2(k^*)$ instead of $f_1(k^*)$,
  we get the non-analytic piece starting at $O(p^3)$, and so on.

  Imagine now that we have only one coupling $H_0$ at our disposal that counts
  at $O(p^0)$. Adjusting this single coupling, one can achieve that
  $\mbox{Re}\,W^{(2)}=O(p)$ if $f_1(k^*)$ is used since the non-analytic
  piece starts at $O(p^3)$. If $f_2(k^*)$ is used, the non-analytic piece starts
  only at $O(p^3)$ and the leading contribution comes from
  the analytic piece at $O(p^2)$, i.e., $\mbox{Re}\,W^{(2)}=O(p^2)$.
  Using $f_3(k^*),\ldots$ in the calculations
  does not lead to the further improvement, since we do not have the
  $H_2$ counterterm
  at our disposal to remove the $O(p^2)$ piece. By the same token, using
  $f_3(k^*)$ should be optimal in case of two constants $H_0,H_2$.
  In this case,  $\mbox{Re}\,W^{(2)}=O(p^4)$ can be achieved.

  Finally, we reiterate that the above discussion should be taken with a grain of salt as it is based on perturbation theory.
    Hence, the counting rules, given above, can provide only a hint about the optimal number of subtractions in the non-perturbative case. We therefore conclude that it is important to numerically check the expectation, based on the above power counting, in non-perturbative calculations. This goal will be accomplished in the next section.

    \section{Numerical test}
    \label{sec:numerics}

In this section, we shall test the approach described above using explicit
nonperturbative calculations. In these calculations, a quantum-mechanical system of three identical
bosons, interacting pairwise through some model potential, will play the
role of an
exact underlying theory. The underlying theory, by definition, does not contain spurious poles. These
appear, when one replaces an exact two-body amplitude in the Faddeev equations
by the effective-range expansion. Thus, one may check, whether the results,
obtained in our scheme, do indeed converge to the known (exact) result, and estimate the rate of this convergence.
We will consider a Yamaguchi potential first and then 
repeat this analysis for a Gauss potential.

\subsection{Yamaguchi model}

As mentioned above, we consider a toy model with three bosons of a mass  $m$, interacting through
the  Yamaguchi potential~\cite{Yamaguchi:1954}, as an exact theory.
This potential is given by:
\begin{align}
    V_Y(p,q)=\lambda \chi(p)\chi(q) \, ,\quad\quad \chi(q)=\frac{\beta^2}{\beta^2+q^2}.
\end{align}
Here, $\lambda$ denotes the strength of the potential, and $\beta$ is related to its
range. To connect the parameters of the  Yamaguchi potential to the scattering length
$a$ and the effective range $r$, we calculate  the two-body scattering amplitude:
\eq
t_Y(p,q,z) =\chi(p)d_Y(E)\chi(q)\, ,\quad\quad
  d_Y(z)=\left[\frac{1}{\lambda}-\int \frac{d^3{\bf q}}{(2\pi)^3}\, \frac{\chi^2(q)}
    {z-E_q} \right]^{-1} \, .
  \en
  The on-shell amplitude takes the form:
  \eq
   t_Y(p,p,E_p)=\chi^2(p)\left[\frac{1}{\lambda}
    -\frac{m\beta^3}{8\pi(p+i\beta)^2}\right]^{-1}\, ,
    \label{twobodyscattering}
\en
where $E_p=p^2/m+i\varepsilon$.
Expanding this amplitude and comparing the result to the effective-range expansion,
we obtain:
\begin{align}
  -\frac{1}{a} =-\frac{\beta}{2}-\frac{4\pi}{\lambda m}\, ,
  \quad\quad
  r =\frac{1}{\beta}-\frac{16\pi}{\lambda\beta^2 m}\,.
\end{align}
In the numerical calculations, the values of $a$ and $r$ are chosen to be equal\footnote{Of course, here we study three
  bosons, so the parameter choice is not directly linked to any real physical problem. However,
  the generalization of our method to the case of the particles with spin is straightforward.}
to the  $np$-triplet scattering parameters
$a=5.4194\text{ fm}$ and $r=1.7563\text{ fm}$. Given this input, one can fix
the parameters of the original Yamaguchi potential. This results in
$\lambda=-0.00013~\text{MeV}^{-2}$ and
$\beta=278.8~\text{MeV}$. The mass $m$ is chosen equal to the proton mass. We shall use these values in the following. For this choice of
parameters, a stable dimer with the mass $M_d=2m-E_d$ emerges.
The binding energy of the dimer, $E_d\simeq 2.22~\mbox{MeV}$, coincides with
that of the deuteron.

In the three-body sector, the model does not contain a three-body force.
This is a valid choice since all integrals are convergent at the upper limit (the parameter
$\beta$ plays a role of the ultraviolet cutoff).
The equation for  the particle-dimer scattering amplitude $M_Y(k,p,E)$ takes
the form (see, e.g., the textbook by Schmid and Ziegelmann~\cite{Schmid}): 
\begin{align}
\begin{split}
M_Y(k,p,E)&=2Z_Y(k,p,E)+\frac{1}{\pi^2}\int dq\,q^2 Z_Y(k,q,E)\tau_Y(q,E)M_Y(q,p,E)\, ,
\label{FaddeevYamaguchi}
\end{split}
\end{align}
where the dimer propagator $\tau_Y(q,E)$ is given by:
\begin{align}
  \label{eq:convention}
\begin{split}
  \tau_Y(q,E)&=d_Y(z)\biggr|_{z=3q^2/(4m)-E-i\varepsilon}
\\[2mm]
  &=\frac{8\pi}{m\beta^3}\frac{(\beta+\gamma)^2(\beta+\sqrt{3q^2/4-mE})^2}{2\beta+\gamma+\sqrt{3q^2/4-mE}}\frac{1}{\gamma-\sqrt{3q^2/4-mE}},
\end{split}
\end{align}
with $\gamma=\sqrt{-m\lambda\beta^3/(8\pi)}-\beta=\sqrt{mE_d}$, and the convention
$E\to E+i\varepsilon$ is implicit everywhere in Eq.~(\ref{eq:convention}).

The one-particle exchange potential $Z_Y(p,q,E)$ in the Yamaguchi model
can be written down in the following form:
\begin{align}\label{eq:ZY}
\begin{split}
  Z_Y(p,q,E)&=\frac{1}{2}\int_{-1}^1 d\cos\theta_{p,q}\frac{\chi({\bf q}+1/2\,{\bf p})
    \chi(-{\bf p}-1/2\,{\bf q})}{E-{\bf p}^2/(2m)-{\bf q}^2/(2m)
    -({\bf p}+{\bf q})^2/(2m)}\\[2mm]
  &=\frac{m}{2}\int_{-1}^1 du\,\frac{\beta^2}{\beta^2+p^2/4+q^2+pqu}\,
  \frac{\beta^2}{\beta^2+p^2+q^2/4+pqu}\\[2mm]
  &\times \frac{1}{mE-p^2-q^2-pqu}\, .
\end{split}
\end{align}
The calculation of the amplitude $M_Y(k,p,E)$ can be carried out by using standard
numerical procedures. We namely use a large momentum cutoff $\Lambda=1500\,\text{MeV}$
to approximate the integral (the presence of the
momentum cutoff is not critical since, as said above, the integral converges even in the
absence of the cutoff).\footnote{Above the breakup threshold, one could
  perform a contour rotation in the integral, or use some other technique, in order
  to circumvent the (integrable) logarithmic singularity of the one-particle exchange potential which hits
  the contour. For simplicity, we will not treat the integrable singularity in any special way. The emerging numerical irregularities are small and do not affect our conclusions.}

In the model, the particle-dimer scattering phase shift
$\delta_Y(p)$ is defined, according to:
\eq
M_Y(p,p,E_p)=-\frac{3m\beta^3}{8\gamma(\beta+\gamma)^3}
\,\,\frac{1}{p\cot\delta_Y(p)-ip}\, .
\en
As already mentioned above, below the dimer breakup threshold $E<0$, the phase shift $\delta_Y(p)$ is real, according to the unitarity. Note also that, in order to ease notations, we
did not choose the same normalization for the amplitudes $M$ and $M_Y$. This does not
cause a problem, since the particle-dimer phase shifts are compared, which are independent
on the normalization chosen.

\subsection{Matching of the EFT framework}

As stated before, our aim is to compare the solution of the Faddeev equation $M_Y(k,p,E)$
with the solution of the Eq.~(\ref{ChangedFaddeev}), where
$W({\bf{p}},{\bf{q}},E)=Z({\bf{p}},{\bf{q}},E)$ is assumed.
In the calculations,
again the hard cutoff is imposed, and two values $\Lambda=250\,\mbox{MeV}$
and $\Lambda=600\,\mbox{MeV}$ are used. Note that, in this case, the cutoff plays a
crucial role as regulator, since the momentum integrals are otherwise divergent.

\begin{sloppypar}
Owing to the initial choice of the parameters, both propagators $\tau_Y$ and
$\tau$ have a pole at the deuteron energy $E_d=k_1^2/m$,
corresponding to $k_1\simeq 46\,\mbox{MeV}$.
For the given choice of parameters,
the quantity $\tau$ exhibits a second, spurious pole at
$k_2\simeq 179\,\mbox{MeV}$ as well, whereas in $M_Y(k,p,E)$, such a pole is absent.

In order to apply our method,
we define the subtracted propagators $\tau_i(k^*)=\tau(k^*)-f_i(k^*)$, where $i$ denotes
the number of subtractions. Thus,
\end{sloppypar}
\begin{align}
\begin{split}
\tau_1(k^*)&=\frac{2(k_2+k_1)/r}{(k_2-k_1)(k^*+k_2)(k^*-k_1)}+\frac{4k_2/r}{(k_2-k_1)k_2^2}\,,\\[2mm]
\tau_2(k^*)&=\frac{2(k_2+k_1)/r}{(k_2-k_1)(k^*+k_2)(k^*-k_1)}+\frac{4k_2/r}{(k_2-k_1)k_2^2}\left\{1+\frac{k^{*2}}{k_2^2}\right\}\,,\\[2mm]
\tau_3(k^*)&=\frac{2(k_2+k_1)/r}{(k_2-k_1)(k^*+k_2)(k^*-k_1)}+\frac{4k_2/r}{(k_2-k_1)k_2^2}\left\{1+\frac{k^{*2}}{k_2^2}+\frac{k^{*4}}{k_2^4}\right\}\, .
\end{split}
\label{eq:Tau}
\end{align}
Note also that, for the remaining (shallow) pole in $\tau_i(k^*)$, the prescription
$k^*\to k^*-i\varepsilon$ is implicit in all above expressions. This corresponds to
$E\to E+i\varepsilon$.

Various approximations, which can be constructed within our approach, differ by
a) order in the effective range expansion and the number of the three-body couplings
$H_0,H_2,\ldots$ used, and b) the number of the retained terms in the expansion in $\tau_i(k^*)$. The calculations are done for leading order (LO), next-to-leading order (NLO) and next-to-next-to-leading order (N$^2$LO) in pionless EFT.
According to the standard power counting in the two- and three-body sectors, the following parameters appear:

\begin{table}[H]
\begin{center}
  \begin{tabular}{c|c|c}
    Order & 2-body parameters & 3-body parameters\\
    \hline
    LO & $a$ & $H_0$ \\
    NLO & $a,\,r$ & $H_0$ \\
    N$^2$LO & $a,\,r$ & $H_0,\,H_2$
  \end{tabular}
\end{center}
\caption{Appearance of 2- and 3-body parameters per order of the EFT power counting.}
\end{table}

Next, we briefly discuss the matching of the low-energy couplings $H_0,H_2$. If there
is only one three-body coupling present, as at  LO and NLO, it is most convenient to
determine it from matching at threshold. For technical reasons, we perform
matching of the particle-dimer scattering phases in two theories $p\cot\delta_Y(p)$ and
$p\cot\delta(p)$ at small, but non-zero value of the momentum
$p=0.001 \text{ MeV}$. When the second coupling $H_2$ is present (N$^2$LO), it would be
natural to match in addition the first derivative of the function $p\cot\delta(p)$ at threshold.
Equivalently, one could match the value of the function $p\cot\delta(p)$ at a some value
of $p$ above threshold. We have opted for the second option, because it
is easier to implement in our numerical algorithm, and have chosen the value
of the second matching momentum $p=10\,\text{ MeV}$, which is still quite close to
threshold.

Below, we shall discuss the matching condition briefly. First,
note that the values of the couplings $H_0$ and $H_2$, in addition
to the cutoff $\Lambda$, depend on the number of the retained terms in the
Taylor-expansion of the spurious pole (this latter dependence is not present in LO,
because there are no spurious poles at this order). Further,
it is seen that the results of the matching for $H_0$ do not depend on, whether
$H_2$ is included or not. This follows from the fact that the contribution from $H_2$
is multiplied by a factor $(mE+\gamma^2)$ (see Eq.~(\ref{eq:ZE})),
which exactly vanishes at the
particle-dimer threshold. This is seen in the Table~\ref{ValuesH}, which
summarizes our final results of the matching of $H_0,H_2$.

\begin{table}[t]
.
\begin{center}  
  \begin{tabular}{cc|cccc}
&$\tau_i$ &$H_0(\Lambda=250)$&$H_2(\Lambda=250)$&$H_0(\Lambda=600)$&$H_2(\Lambda=600)$\\ \hline
    LO&  &-5.30&&0.40&\\
\hline
    \multirow{3}{*}{NLO \& N$^2$LO}
&$\tau_1$&-0.82&0.25&0.86&-8.20\\  
&$\tau_2$&-1.17&0.63&-1.11&2.01\\  
&$\tau_3$&-1.31&0.84&7.41&2223.\\
 \end{tabular}
  \end{center}
  \caption{The three-body couplings $H_0$ and $H_2$ for the
    different values of the cutoff $\Lambda$, and different number of subtractions
    in the propagator $\tau(k)$ (no subtraction is needed at LO). All quantities are given  in MeV units. The values of $H_0$ are the same at NLO and N$^2$LO, whereas $H_2=0$ at NLO.}
\label{ValuesH}
\end{table}

\subsection{Numerical results for the phase shift}

\begin{figure}[htb]
\subfigure{\includegraphics[width=0.49\linewidth]{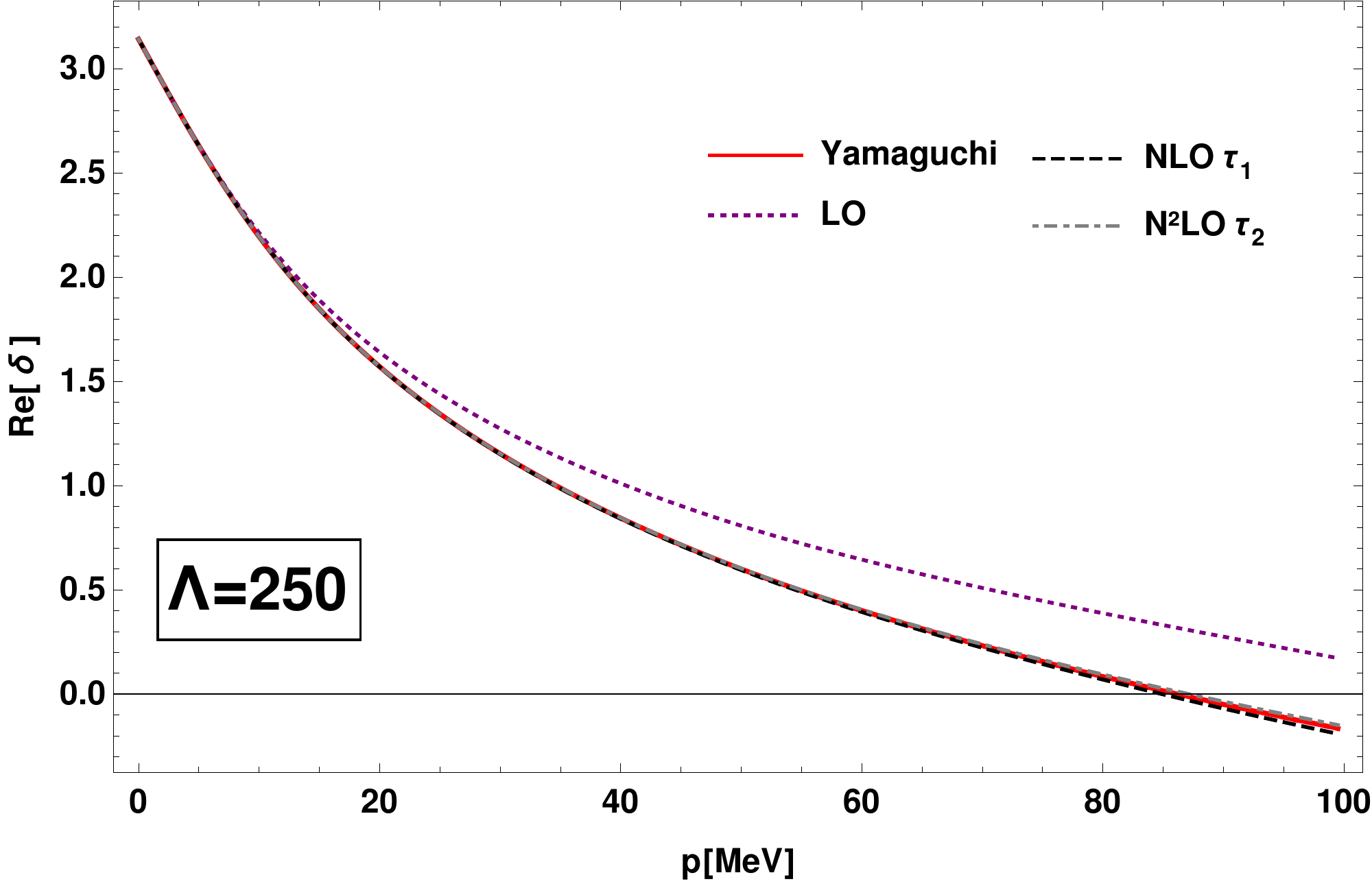}}
\hfill
\subfigure{\includegraphics[width=0.49\linewidth]{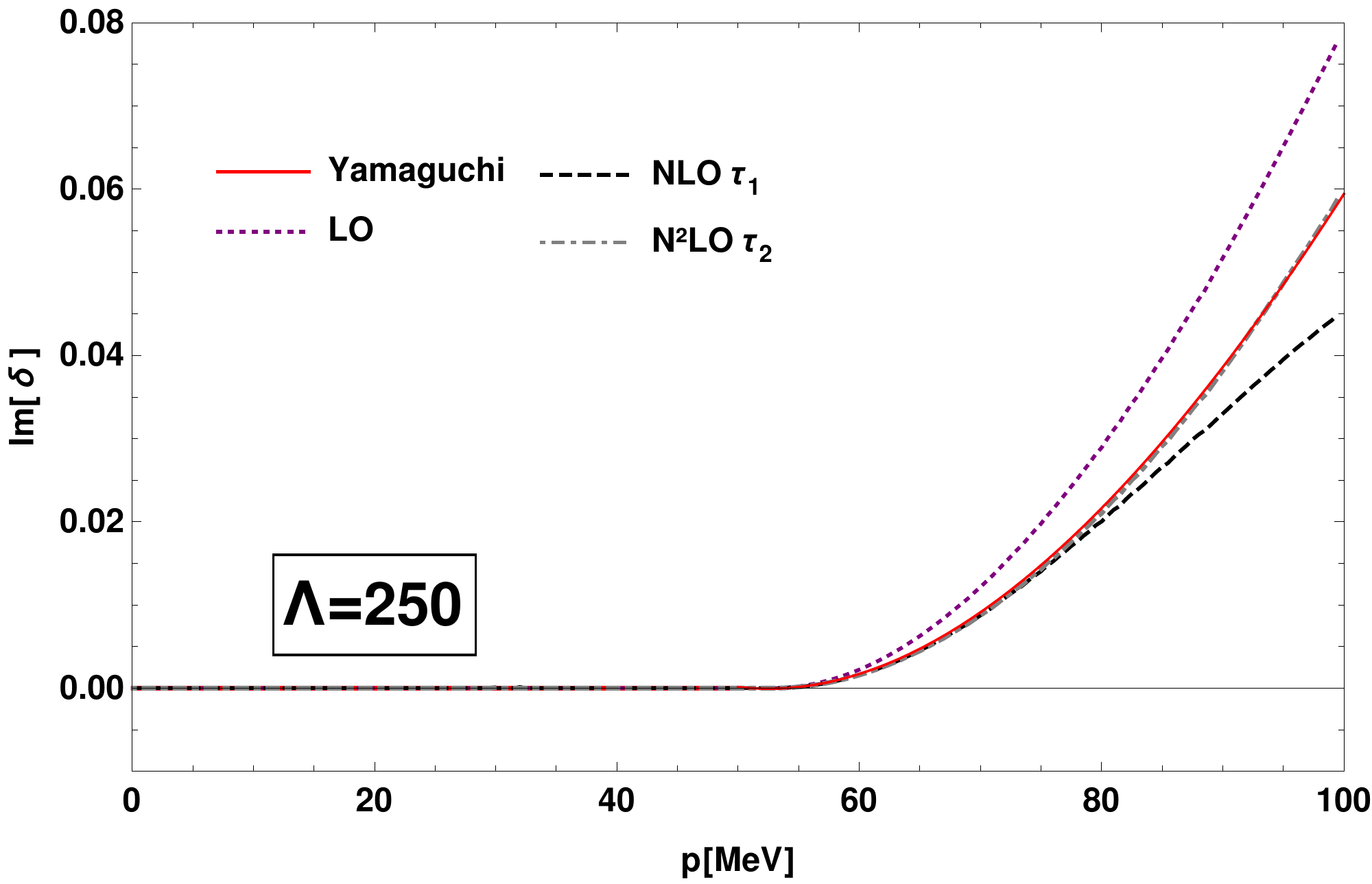}}
\caption{Numerical results for real (left) and imaginary (right) part of the particle-dimer phase shift $\delta$ for the Yamaguchi model. Red line: the result obtained in the Yamaguchi model; in purple dotted: the LO result; in black dashed: the NLO result for $\tau_1$; in gray dot-dashed: the N$^2$LO result for $\tau_2$. For the real part the NLO and N$^2$LO results are on top of the Yamaguchi model. For the imaginary part the N$^2$LO results are on top of the model. The cut-off was set to the value $\Lambda=250~\text{MeV}$.}
\label{YamaguchiDelta}
\end{figure}

To begin with, we calculate the particle-dimer scattering phase shift $\delta$ in the toy model with the Yamaguchi potential, and in the effective theory, amended by our prescription for treating the spurious poles. As mentioned above, we have in fact to deal with two different expansions: the EFT expansion (i.e., including more derivative terms in the Lagrangian that are accompanied with the independent couplings), and the Taylor-expansion of the spurious pole. The convergence of these expansions need to be investigated separately.
Since it turns out to be the most efficient choice, we use the subtracted propagators $\tau_1(k^*)$ and $\tau_2(k^*)$ in the calculations at NLO and N$^2$LO respectively. Remember that at LO, no subtraction is needed.
Note also that this choice differs from our perturbative estimate in
Sec.~\ref{sec:formalism} by one order.
The other possible choices of $\tau_i(k^*)$ at NLO and N$^2$LO,
including the one based on perturbation theory, are discussed below.
The real part of the results of these calculations are shown in the left part of Fig.~\ref{YamaguchiDelta}. It is seen that LO is precise only at small momenta, whereas NLO can describe data at much higher values of $p$. The situation further improves at N$^2$LO, albeit this improvement is very small (practically not visible by a bare eye). In the left part of Fig: \ref{YamaguchiDelta} the imaginary part of $\delta$ is shown. It can be seen that the NLO and N$^2$LO results describe the model better than LO, while the N$^2$LO results are clearly improved compared to NLO.
The errors of the EFT calculation for $p>1/a$ can be estimated as
$(p/\Lambda)^{n+1}$ at N$^n$LO. A more detailed evaluation of the EFT
errors is presented in the discussion of possible choices for $\tau_i(k^*)$
below.

Up to now, everything follows the standard EFT pattern. However, in order to answer the question, whether a systematic improvement is achieved in higher orders, as well as to address the subtraction of the spurious pole, a more elaborate study of the problem is necessary. To this end, it is convenient to use the so-called Lepage plots, which will be considered below.

\subsection{Lepage plots and consistency assessment}

Lepage \cite{Lepage:1997} has proposed a method, which allows one to check,
how well the data are described by an EFT. The method makes use of certain
double-logarithmic plots, known as the Lepage plots.
Grie{\ss}hammer~\cite{Griesshammer:2020} has suggested
to verify the internal consistency of an EFT along the similar pattern.
In the following,
we shall adapt these methods for the problem we are working on.

Let us consider an EFT, describing the fundamental theory up to order $n$.
The corrections are of the order $[(k_{typ},p)/\Lambda_b]^{n+1}$, where
$k_{typ}\sim 1/a$ is a typical momentum in the reaction and $\Lambda_b$ is
the breakdown scale of an EFT. For an arbitrary observable, and, in particular,
for  the three-body phase-shift $p \cot (\delta)$, we have:
\begin{align}
  \frac{p \cot (\delta_{Data})-p \cot (\delta_{EFT})}{p \cot (\delta_{Data})}& = c \left(\frac{(k_{typ},p)}{\Lambda_b}\right)^{n+1-\eta}+\cdots\, .
\end{align}
This means that
\begin{align}
 \ln\left[\frac{p \cot (\delta_{Data})-p \cot (\delta_{EFT})}{p \cot (\delta_{Data})}\right] & \approx c'+ (n+1-\eta) \, \ln\left[\frac{p}{\Lambda_b}\right]=c''+ (n+1-\eta) \, \ln\left[p\right]\, .
\end{align}
Here, $c$, $c'$ and $c''$ stand for some constants. The quantity $\eta$ describes the corrections due to the denominator. It is also assumed that $k_{typ}\ll p$, this is discussed below. Hence, the slope in a double-logarithmic plot gives the order $n$ of the neglected term.
To determine this slope, a linear function can be fitted to the numerical results.

\begin{figure}[htb]
\subfigure{\includegraphics[width=0.49\linewidth]{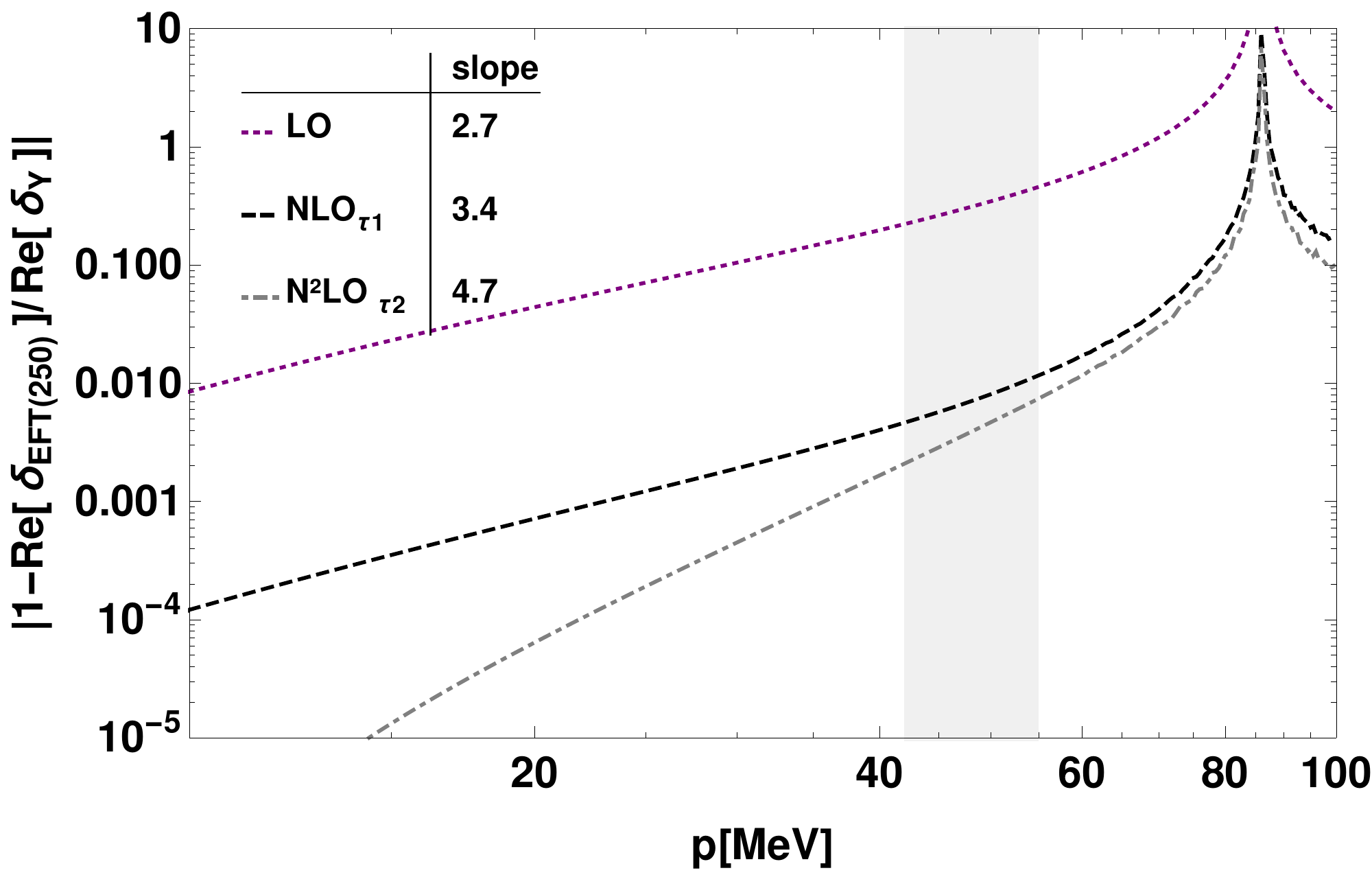}}
\hfill
\subfigure{\includegraphics[width=0.49\linewidth]{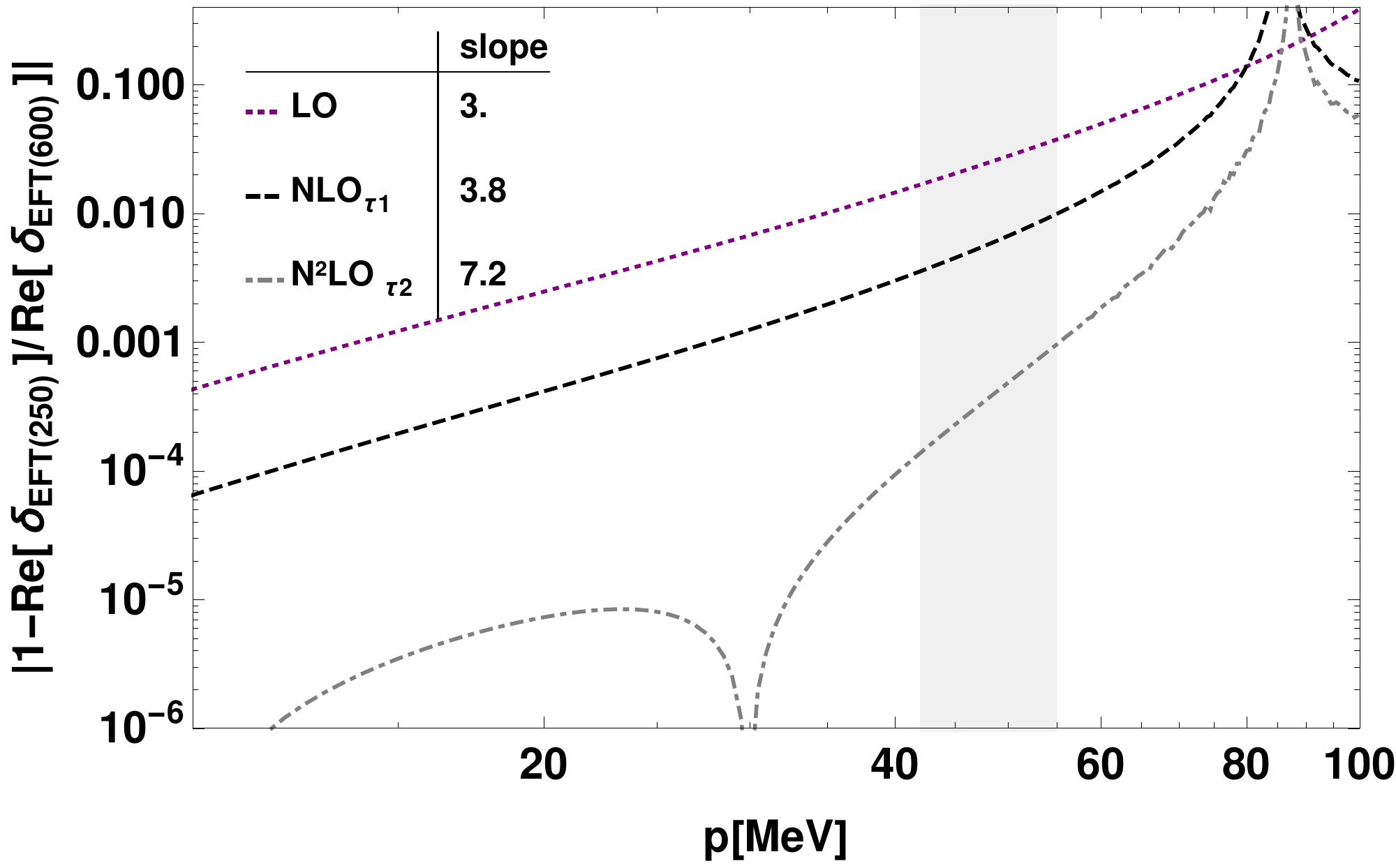}}
\caption{Lepage plot (left) and consistency assessment (right) for the particle-dimer phase shift in the Yamaguchi model. The ``window of opportunity'' is chosen to be between 42 MeV and 55 MeV for all orders (gray shaded region). The spike (zero of $\delta$ (Fig. \ref{YamaguchiDelta})) around 80 MeV limits us to low energy regions. Note that the LO result does not predict this zero, therefore the spike is not visible in the consistency assessment at LO, for the Lepage plot the results are divided by the Yamaguchi results, therefore the spike can be seen at all orders. As expected the slope is increasing by approximately one order by order. The deviant value for N$^2$LO $\tau_2$ is due to the accidental zero around 30 MeV (change in the sign), compare to \cite{Griesshammer:2020}.}
\label{YamaguchiDeltaLepage}
\end{figure}

Further, one may check the internal consistency of
an EFT without comparing to data at all~\cite{Griesshammer:2020}.
Instead, one can compare the results of
calculations within the same EFT, at two different values of the ultraviolet cutoff
$\Lambda_1$ and $\Lambda_2$,  
\begin{align}
\begin{split}
\frac{p \cot (\delta_{EFT(\Lambda_2)})-p \cot (\delta_{EFT(\Lambda_1)})}{p \cot (\delta_{EFT(\Lambda_2)})}&= c(\Lambda_1,\Lambda_2,k_{typ},p,\Lambda_b) \left(\frac{(k_{typ},p)}{\Lambda_b}\right)^{n+1-\eta}+...
\end{split}
\label{slope}
\end{align}
Here $c(\Lambda_1,\Lambda_2,k_{typ},p,\Lambda_b)$ is a slowly varying function of $k_{typ}$ and $p$.
Further, the parameter $\eta$ describes the dependence
of $p \cot (\delta_{EFT(\Lambda_2)})$ on $p$ at LO and will be determined from the fit at
the LO. The slope in a double logarithmic plot is, approximately, $n+1-\eta$. Note that the $\eta$ in the consistency assessment and the Lepage plots
does not have to be the same.
Since $k_{typ}$ is not uniquely determined and the double expansion in
${k_{typ}}/{\Lambda_b}$ and ${p}/{\Lambda_b}$ complicates the analysis,
it is very useful to stick to the region,
 \eq
 k_{typ}\ll p\ll\Lambda_b\sim\Lambda\,.
 \label{Window}
 \en
Moreover, we choose the cutoff $\Lambda$ of the order of the breakdown scale
 $\Lambda_b$ to simplify the analysis.
 In this region, termed as the ``window of opportunity'',
 the dependence on $k_{typ}$ should disappear (recall that, in our case,
 $k_{typ}=1/a$). On the other hand,
 one cannot use too large values of the variable $p$, of the order of the hard scale $M_{high}$ of the theory,
 determined by the effective range and/or ultraviolet cutoff.
  Hence, ensuring that one can reliably
 determine the slopes from the fits in the ``window of opportunity'' is a non-trivial exercise. For example in Fig.~\ref{YamaguchiDeltaLepage} we see that around $80$ MeV a spike appears. This is due to $Re[\delta]=0$ (compare to Fig.~\ref{YamaguchiDelta}) in the denominator. This spike will change the slope in this region, so the ``window of opportunity'' is restricted to be below this. With this we choose the window between 42 MeV and 55 MeV for the $\delta$-slopes.

\begin{table}[htb]
    \begin{minipage}{.5\linewidth}
      \centering
      \begin{tabular}{l|lll}
slope fit& LO & NLO & N$^2$LO\\  \hline
no sub. & {2.7} &   &\\
$\tau_1$ &&  3.4 & 4.4\\
$\tau_2$&  &  3.6 & 4.7\\
$\tau_3$&  &  3.6 & 5.0\\
  \end{tabular}
    \end{minipage}%
    \begin{minipage}{.5\linewidth}
      \centering
       \begin{tabular}{l|lll}
slope fit & LO & NLO & N$^2$LO\\ \hline
no sub. &3.0&&\\
  $\tau_1$ &&  3.8 & 5.3\\
 $\tau_2$&  &  4.0 & 7.2*\\
 $\tau_3$&  &  4.0 & 4.9\\
  \end{tabular}
    \end{minipage} 
    \caption{Results for the slopes for the particle-dimer phase shifts $\delta$ fitted in the ``window of opportunity'' for the Yamaguchi model. The uncertainty in the slopes is about 10\%.
      Left for the Lepage plot right for the consistency assessment. The value with the asterisk * is unnaturally large due to a accidental zero, compare to Fig. \ref{YamaguchiDeltaLepage} (right).}
    \label{SlopeYamaguchiLepage}
\end{table}

 We start with the slope fits, using the subtracted propagators $\tau_1(k^*)$ and $\tau_2(k^*)$ for NLO and N$^2$LO respectively, we analyze the results for the real part of the particle-dimer phase shift $Re[\delta]$. The plots are shown in Fig.~\ref{YamaguchiDeltaLepage}. The slopes are increasing order by order as expected, for the Lepage-plots (left) as well as for the consistency assessment (right). The exact value of the increase should be one per order. In the left part of Table \ref{SlopeYamaguchiLepage} the slopes are shown for the Lepage plot, in the right part the slopes for the consistency assessment.   
 The slopes for other choices of $\tau_i(k^*)$ are also included.
 By varying the ``window of opportunity'' slightly, we estimate the uncertainty in determination of these slopes
 from the fit at about 10\%.
 Note that the value for $\eta$ can not be predicted \cite{Griesshammer:2020}, it is determined by the slope for the LO results.\footnote{This is also the reason for the values for $\delta$ being different from the values for $k\cot\delta$ seen in Table \ref{Slope}.} It can be seen, that all results agree with the predicted increase approximately. Note that the result for N$^2$LO $\tau_2(k^*)$ in the consistency assessment is an exception due to the accidental zero (compare to the discussion in Fig. \ref{YamaguchiDeltaLepage}). The values for N$^2$LO using $\tau_1(k^*)$ or $\tau_3(k^*)$ are close to the expected value of 5, the corresponding graphs do not exhibit the accidental zero.

Taking into account the 10\% uncertainty in the determination of
the slope,
the results in Table \ref{SlopeYamaguchiLepage} show that using
$\tau_2$ and $\tau_3$ at NLO leads to no significant improvement of the slope
compared to $\tau_1$. This provides a justification for our choice of using
$\tau_1$ at NLO. Since we have one more constant, $H_2$, at our disposal
at N$^2$LO, one more subtraction can be accommodated. This motivates our use of
$\tau_2$ instead of $\tau_1$ at N$^2$LO despite the insignificant improvement
in the slope.

\begin{table}[htb]
\begin{center}
  \begin{tabular}{l|lll}
slope & LO & NLO & N$^2$LO\\ \hline
fit~\cite{Griesshammer:2020} & 1.9 & 2.9 & 4.8\\
our fit, no sub. & 1.8 &  & \\
our fit, $\tau_1$ &  &  2.8 & 4.6\\
our fit, $\tau_2$&  & 2.9 & 6.1*\\
our fit, $\tau_3$&  & 2.8 & 3.6\\
  \end{tabular}
  \end{center}
\caption{Slope fits for $k \cot \delta$ in the consistency assessment for
  the Yamaguchi model. The ``window of opportunity'' was chosen between $42$ MeV and $55$ MeV.
  The uncertainty in the slopes is about 10\%.
  Shown are the fits to the results for $Re[p\cot \delta]$.
The value with the asterisk * is unnaturally large due to a accidental zero.}
\label{Slope}
\end{table}

Additionally, we have repeated the same analysis for $k \cot \delta$ instead of
the phase shift $\delta$ (the observable considered in Ref.~\cite{Griesshammer:2020}).
The extracted slopes in Table~\ref{Slope}
are again consistent with our choice
$\tau_1(k^*)$ and $\tau_2(k^*)$ in the calculations at NLO and N$^2$LO,
respectively.

To summarize, solving the scattering equation for the particle-dimer amplitude in EFT,
while treating the spurious pole as proposed above, we have explicitly demonstrated
that the numerical solution systematically converges to the exact result, obtained in the
Yamaguchi model, which does not contain spurious poles. Moreover, the pattern of this
convergence, in general, follows the theoretical predictions. Hence, the theoretical
construction of Sect.~\ref{sec:formalism} has been verified.

\subsection{Order of the subtraction polynomial and numerical results}

\begin{figure}[htb]
\subfigure{\includegraphics[width=0.49\linewidth]{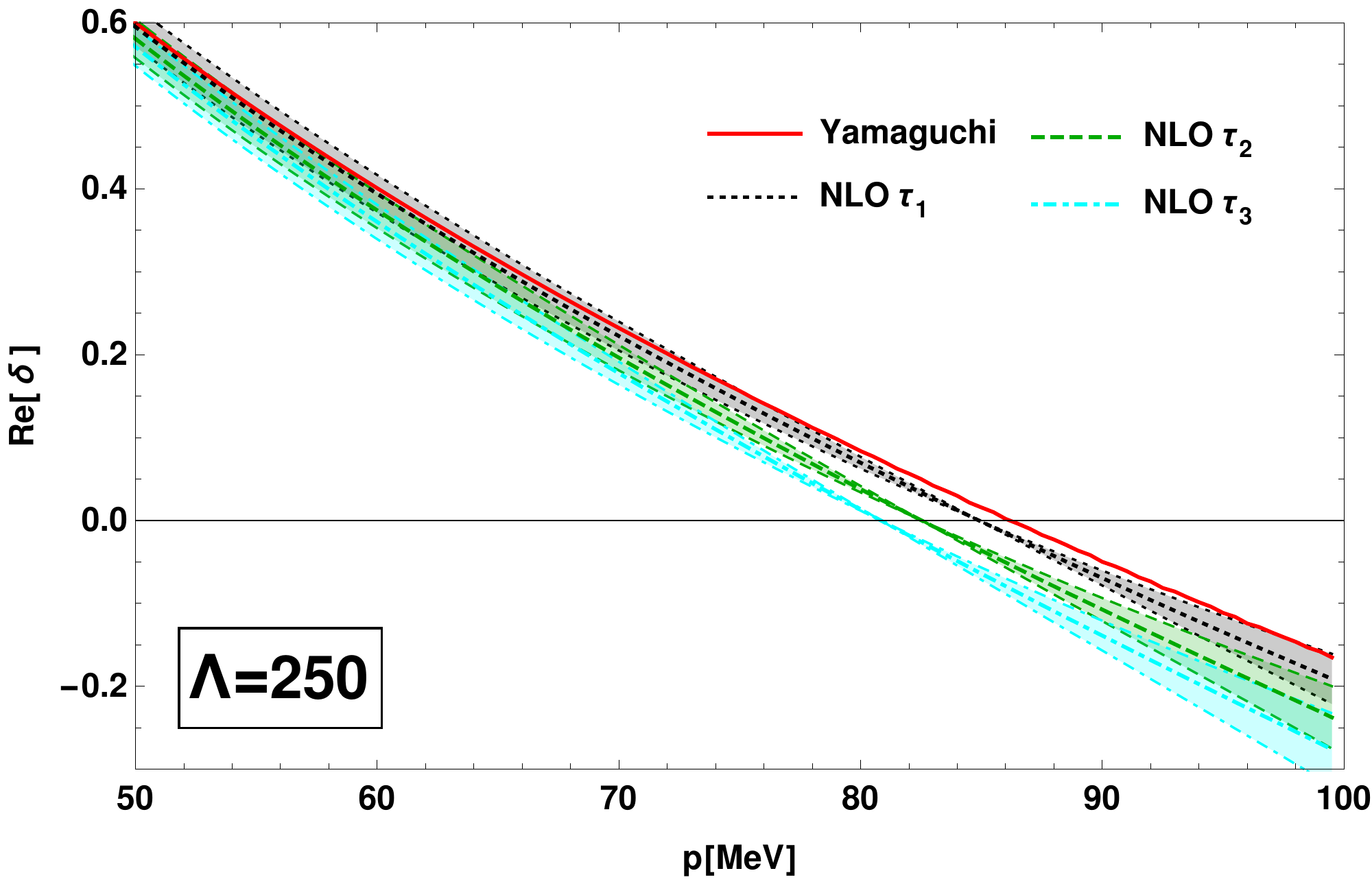}}
\hfill
\subfigure{\includegraphics[width=0.49\linewidth]{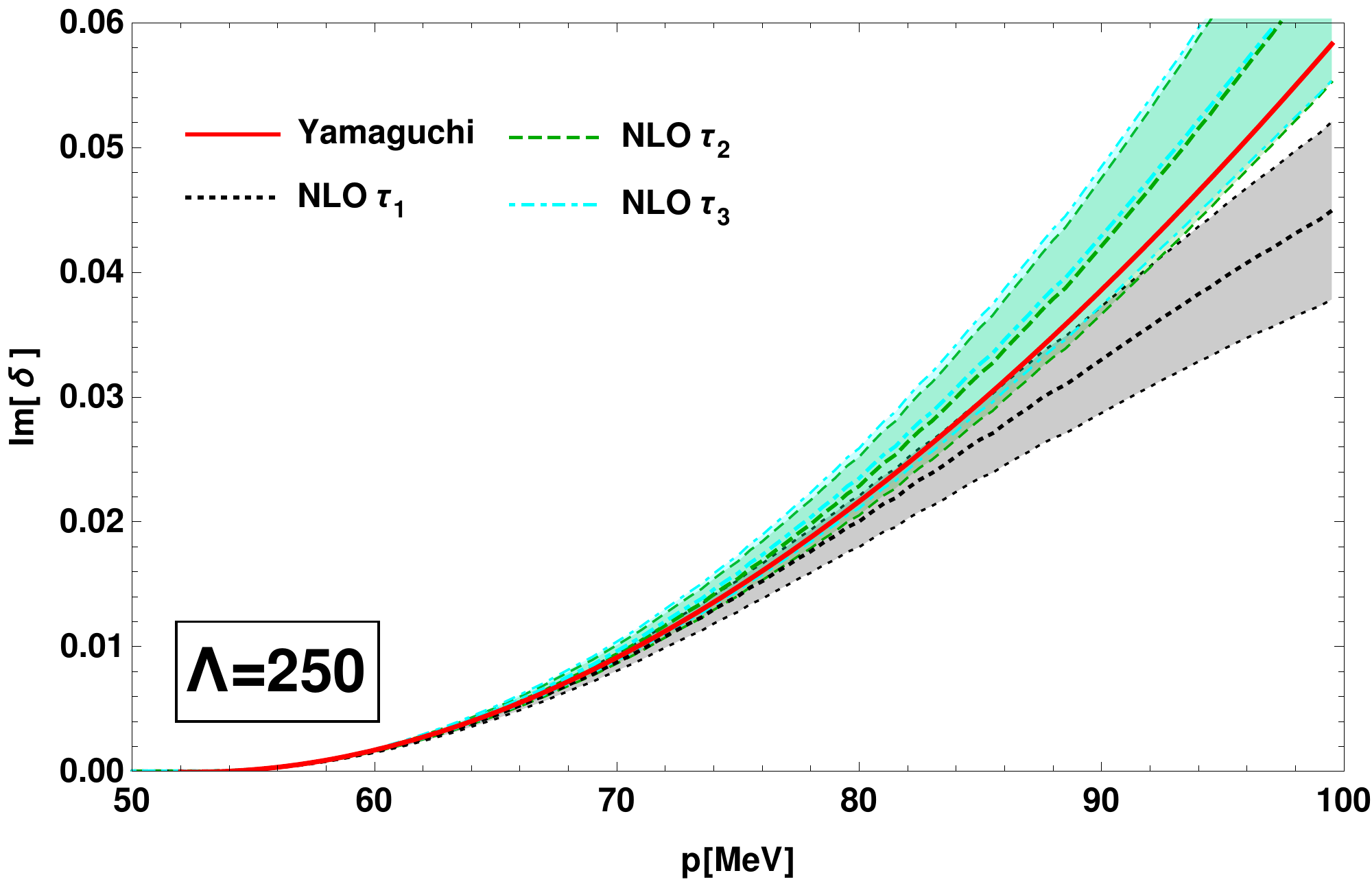}}
\hfill
\caption{Real (left) and imaginary (right) part of the particle-dimer phase shift $\delta$ calculated for the Yamaguchi model and the EFT at NLO for different choices of number of subtractions in the propagator $\tau_i(k^*)$. The uncertainty bands are estimated by a naive power-counting of the EFT error, given by $Re[\delta]\,\left(p/\Lambda\right)^2$ and $Im[\delta]\,\left(p/\Lambda\right)^2$.
}
\label{YamaguchiDeltaNLO}
\end{figure}

In the last subsection we have focused on the consistency and model description of the EFT-expansion. We have provided some evidence for our choice
$\tau_1(k^*)$ and $\tau_2(k^*)$ in the calculations at NLO and N$^2$LO,
respectively, based on the behavior of the slopes in Lepage and consistency
plots. In this subsection the optimal order of the subtraction polynomial is investigated further, providing additional justification for the choice done earlier. Namely, the numerical calculations discussed in the last sections are repeated for different orders, which means different choices of $\tau_i(k^*)$ as defined in equation (\ref{eq:Tau}). In the left part of Fig. \ref{YamaguchiDeltaNLO} the EFT results at NLO for different $\tau_i(k^*)$ are compared with the Yamaguchi model for the real part of $\delta$. It becomes clear that $\tau_1(k^*)$ describes the model the best. Further subtractions do not improve the reproduction of the model, they actually make it worse.
This means one subtraction seems to be optimal. The right part of Fig. \ref{YamaguchiDeltaNLO} shows the corresponding imaginary part, here a tiny improvement from $\tau_1(k^*)$ to $\tau_2(k^*)$ is visible. However, this is only true for very large values of the momentum $p$ and the improvement is well below the expected EFT accuracy. The results for $\tau_1(k^*)$ agree with the Yamaguchi model everywhere within the EFT uncertainty. 
There is no improvement from $\tau_2(k^*)$ to $\tau_3(k^*)$ at all. To conclude, at NLO the phase shift can be described most accurately using $\tau_1(k^*)$.
This choice is consistent with the slopes for the Lepage plots and the consistency assessments shown in Table \ref{SlopeYamaguchiLepage} and Table \ref{Slope}, where all slopes for NLO agree within $10\%$. Therefore we choose the minimal amount of subtractions in the following, which means using $\tau_1(k^*)$ at NLO.

\begin{figure}[htb]
\subfigure{\includegraphics[width=0.49\linewidth]{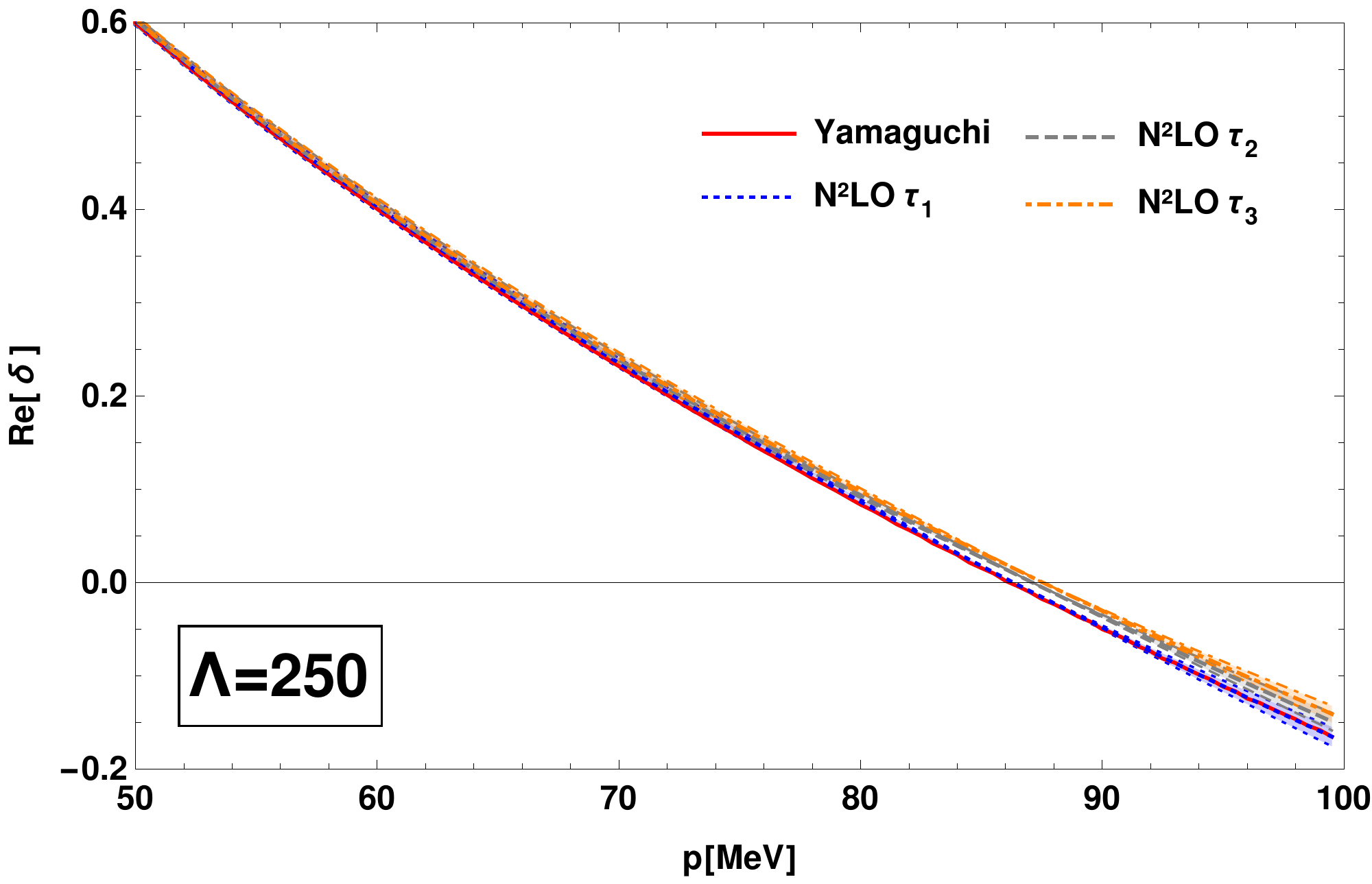}}
\hfill
\subfigure{\includegraphics[width=0.49\linewidth]{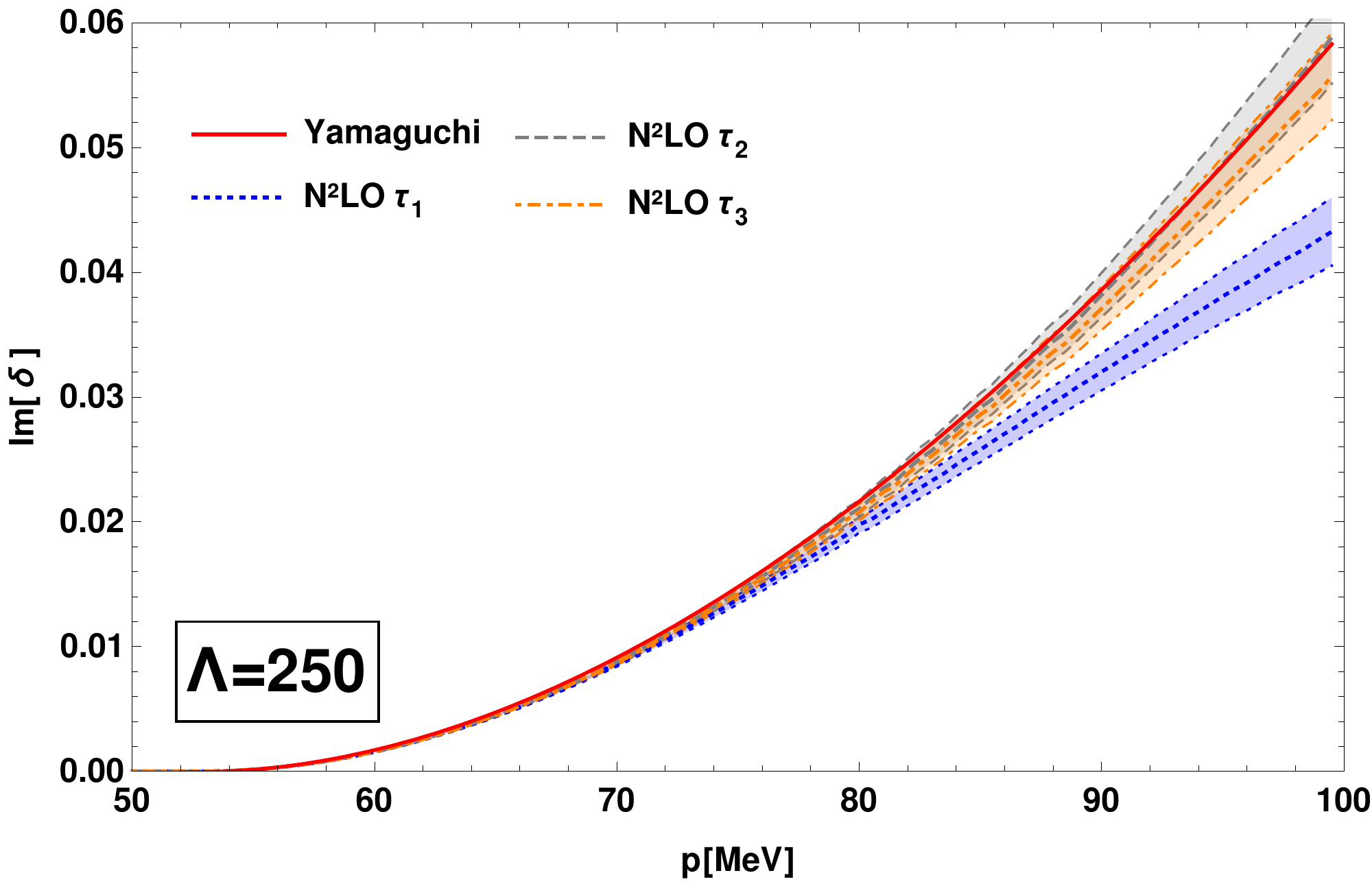}}
\hfill
\caption{Real (left) and imaginary (right) part of the particle-dimer phase shift $\delta$ calculated for the Yamaguchi model and the EFT at N$^2$LO for different choices of number of subtractions in the propagator $\tau_i(k^*)$. The uncertainty bands are estimated by a naive power-counting of the EFT error, given by $Re[\delta]\,\left(p/\Lambda\right)^3$ and $Im[\delta]\,\left(p/\Lambda\right)^3$.
}
\label{YamaguchiDeltaN2LO}
\end{figure}

In Fig. \ref{YamaguchiDeltaN2LO} the phase shift is shown for N$^2$LO. The results appear to be similar to the situation for NLO, the real part is described the best by $\tau_1(k^*)$, while the difference between $\tau_{2,3}(k^*)$ and the model is larger. However, the effect is small in N$^2$LO, as all three choices of $\tau_i(k^*)$ agree within a power-counting estimation of the EFT uncertainty in the wide interval including the opportunity window.\footnote{We estimate our {\em relative}
  error as $(p/\Lambda)^n$, with $n=2,3$ at NLO and N$^2$LO,
  respectively. The {\em absolute} error in the real part then vanishes
  at the energy where $Re[\delta]=0$, thus indicating on the
  natural limitations of such a crude estimate. In fact, one expects
that the absolute error does not change much in the interval considered.}
For the imaginary part, however, the increase from $\tau_1(k^*)$ to $\tau_2(k^*)$ is large. The reproduction of the imaginary part of the model is better for $\tau_2(k^*)$ than $\tau_1(k^*)$. But again from $\tau_2(k^*)$ to $\tau_3(k^*)$ no improvement can be seen. Since the differences for the real part are not significant and the imaginary part clearly indicates to use $\tau_2(k^*)$, we choose $\tau_2(k^*)$ for the N$^2$LO calculations.

\subsection{Yamaguchi model with different $r$}

\begin{table}[htb]
\begin{center}  
  \begin{tabular}{cc|cccc}
&$\tau_i$ &$H_0(\Lambda=250)$&$H_2(\Lambda=250)$&$H_0(\Lambda=600)$&$H_2(\Lambda=600)$\\ \hline
    LO&  &-14.31&&1.46&\\
\hline
    \multirow{3}{*}{NLO \& N$^2$LO}
&$\tau_1$&3.79&11.12&-1.14&-0.42\\  
&$\tau_2$&3.03&10.02&-1.55&0.76\\ 
&$\tau_3$&2.90&9.92&-1.72&1.52\\  
 \end{tabular}
  \end{center}
\caption{The three-body couplings $H_0$ and $H_2$ for the
  Yamaguchi model with $r=0.8768$~fm for
    different values of the cutoff $\Lambda$, and different number of subtractions
    in the propagator $\tau(k)$ (no subtraction is needed at LO). All quantities are given  in MeV units. The values of $H_0$ are the same at NLO and N$^2$LO, whereas $H_2=0$ at NLO. The values for $H_0$ are fine-tuned at $p=0.5 \text{ MeV}$ and $H_2$ at $p=20 \text{ MeV}$.}
\label{ValuesHShifted}
\end{table}

\begin{figure}[htb]
\center
\includegraphics[width=0.6\linewidth]{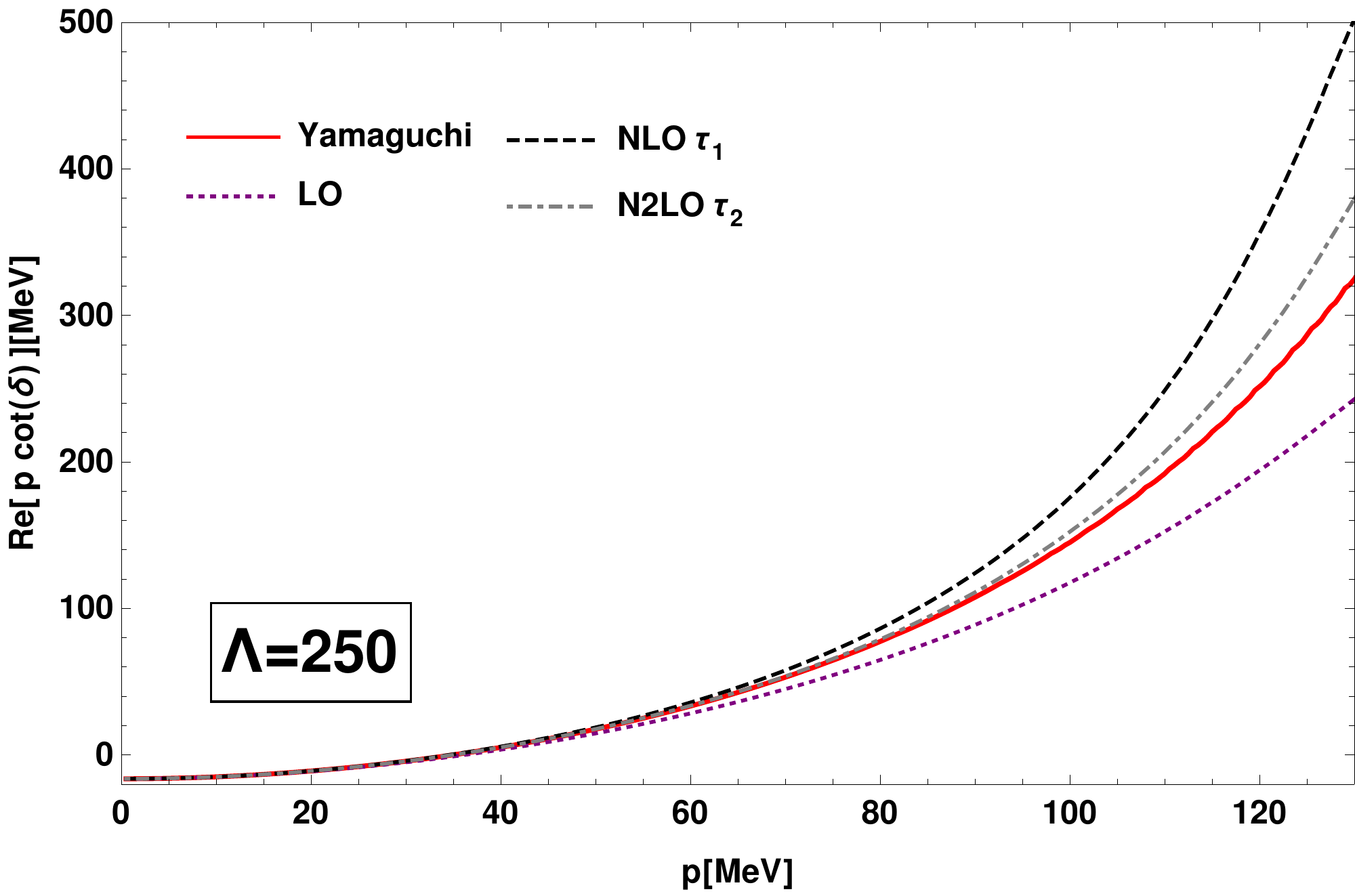}
\caption{Numerical results for the real part of the quantity $p\cot\delta(p)$ for the
  Yamaguchi model with $r=0.8768$~fm. Red line:
  the result obtained in the model with Yamaguchi potential;
  in purple dotted: the LO result; in black dashed: the NLO result for $\tau_1$; in gray dot-dashed: the N$^2$LO result for $\tau_2$. The cut-off was set to the value $\Lambda=250~\text{MeV}$.}
\label{YamaguchikcotdShift}
\end{figure}

The results in the last chapters show a zero for the phase shift $\delta=0$ around $p=80\text{ MeV}$ for the Yamaguchi model. As discussed above this makes the determination of the slope difficult, and limits the ``window of opportunity'' to low energies. To test our method for higher values of the window, a different choice for the effective range is investigated. We choose $r'=0.8768 \text{ fm}(=0.5 r)$ and the same $a=5.4194 \text{ fm}$ as before. This moves the zero outside the considered energy region. Besides the unphysical pole is shifted to $k_2=410.149 \text{ MeV}$. The corresponding Yamaguchi parameters are given by $\lambda=-0.000049 \text{ MeV}^{-2}$ and $\beta=622.5$ MeV. The values for the three-body forces are summarized in Table \ref{ValuesHShifted}. The results for the quantity $k\cot \delta$ are shown in Fig.~\ref{YamaguchikcotdShift}. The pole around $p=80$ MeV is not present anymore. Everything else follows the pattern described for the Yamaguchi model for $r=1.7536$ fm. The description becomes better with increasing orders of the EFT.

\begin{figure}[htb]
\center
\includegraphics[width=0.7\linewidth]{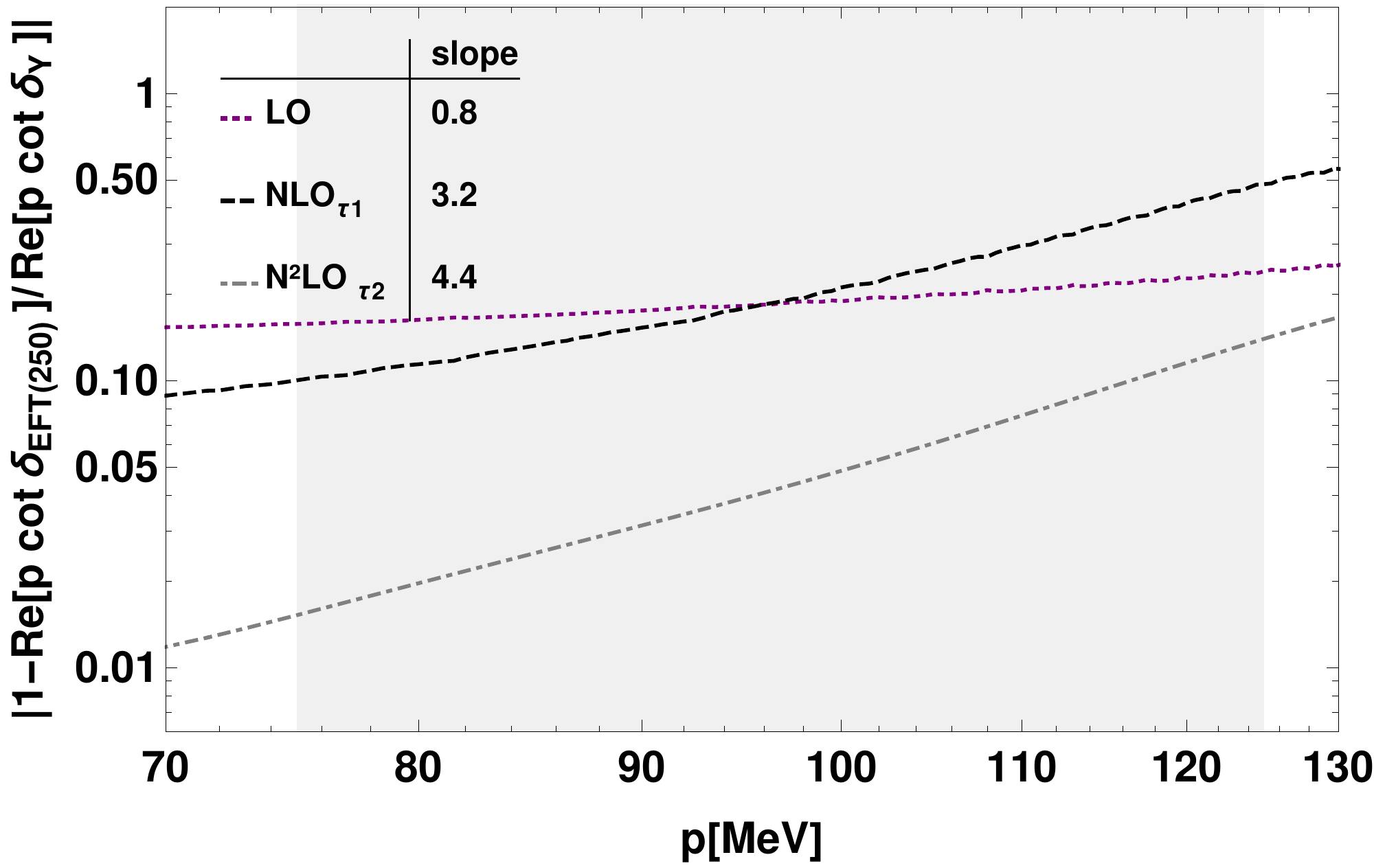}
\caption{Lepage plot compared with the Yamaguchi model
  with $r=0.8768$~fm for the quantity $p\cot\delta$. The ``window of opportunity'' (shaded in gray) is chosen to be between 75 MeV and 125 MeV for all orders.}
\label{YamaguchkcotdShiftLepage}
\end{figure}

Since the spike is shifted we are able to choose higher values for the ``window of opportunity''. We choose it to be between $75\text{ MeV}$ and $125 \text{ MeV}$. In Fig.~\ref{YamaguchkcotdShiftLepage} the Lepage plot is shown. The different orders of the EFT can clearly be distinguished. The slopes increase order by order (see Table \ref{SlopeYamaguchiShift}),
but the increase in the Lepage plot from LO to NLO is slightly larger than expected. With the spike shifted, the slopes are stable under small changes to the ``window of opportunity''.
The general pattern of the method behaves as expected and strengthens the assumptions.

\begin{table}[htb]
    \begin{minipage}{.5\linewidth}
      \centering
      \begin{tabular}{l|lll}
slope fit& LO & NLO & N$^2$LO\\  \hline
no sub. & {0.8} &   &\\
$\tau_1$ &&  3.2 & 4.6\\
$\tau_2$&  &  3.0 & 4.4\\
$\tau_2$&  &  3.1 & 4.4\\
  \end{tabular}
    \end{minipage}%
    \begin{minipage}{.5\linewidth}
      \centering
       \begin{tabular}{l|lll}
slope fit & LO & NLO & N$^2$LO\\  \hline
no sub. &2.2&&\\
  $\tau_1$ &&  3.0 & 4.4\\
 $\tau_2$&  &  3.1 & 4.3\\
 $\tau_2$&  &  3.1 & 4.3\\
  \end{tabular}
    \end{minipage} 
     \caption{Results for the slopes for the quantity $p\cot\delta$ for the
  Yamaguchi model with $r=0.8768$~fm fitted in the ``window of opportunity''. Left for the Lepage plot right for the consistency assessment.}
    \label{SlopeYamaguchiShift}
\end{table}

\subsection{Gauss model}

To further check our results, we perform the same analysis for an additional model potential, namely a Gauss potential. For the Gauss model the  regulator is given by
\begin{align}
\chi(p)=e^{-p^2/\lambda_G^2}
\end{align}
Similar to the Yamaguchi model (compare to equation (\ref{twobodyscattering})), this leads to (for $E<0$)
\begin{align}
\begin{split}
d_G(E)^{-1}&=2\pi^2\biggl[\sqrt{mE_d}\exp\left(\frac{2mE_d}{\lambda_G^2}\right)\text{erfc}\left(\frac{\sqrt{2mE_d}}{\lambda_G}\right)-\sqrt{-E}\exp\left(\frac{-2E}{\lambda_G^2}\right)\text{erfc}\left(\frac{\sqrt{-2E}}{\lambda_G}\right)\biggr]\\
&=2\pi^2\sqrt{mE_d}\exp\left(\frac{2mE_d}{\lambda_G^2}\right)\text{erfc}\left(\frac{\sqrt{2mE_d}}{\lambda_G}\right)+2\pi^2 ip-\frac{4\sqrt{2}\pi^{3/2}}{\lambda_G}p^2+O\left(p^3\right)
\label{Gaussd}
\end{split}
\end{align}
To connect the EFT parameters to the model, we choose $\lambda_G$ to fulfill
\begin{align}
  \frac{1}{a} =\sqrt{mE_d}\exp\left(\frac{2mE_d}{\lambda_G^2}\right)\text{erfc}\left(\frac{\sqrt{2mE_d}}{\lambda_G}\right)\, ,
  \quad\quad
  \frac{r}{2} =\frac{4\sqrt{2}\pi^{3/2}}{\lambda_G2\pi^2}\,.
\end{align}
For the effective range $a=5.4194$ fm and $r=1.7536$ fm, this results in $\lambda_G=359.134$ MeV. In equation (\ref{Gaussd}) the parameter $E_d$ is a input value to the Gauss model, so the two parameters $\lambda_G$ and $E_d$ are described by the two parameters $a$ and $r$ in the EFT. However $E_d$ is equivalent to the position of the root of $d_G^{-1}(E)$ and therefore is the value of a two-body bound state. So for the chosen values of $a$ and $r$ it can be identified with the binding energy of the deuteron, $E_d\approx 2.22 \text{ MeV}$. 
The dimer-propagator $\tau(q,E)$ is given by
\begin{align}
\tau(q,E)=d_G(z)\big{|}_{z=3q^2/(4m)-E-i\epsilon}
\end{align}
In the numerical calculations we use the un-expanded equation for $d_G(z)$ (first line of equation (\ref{Gaussd})).
The one-particle exchange
in the Gauss model, $Z_G(p,q,E)$, is given by a formula similar
to Eq.~(\ref{eq:ZY}). 
To avoid numerical difficulties regarding the poles of the
angular integral, this is calculated partly analytically and partly numerically. For more details regarding this see the appendix.

The values for the three-body forces are fine-tuned to reproduce the Gauss results at $p=0.001$ MeV for $H_0$ and at $p=10$ MeV for $H_2$. The values can be seen in Table \ref{ValuesHGauss}.

\begin{table}[htb]

\begin{center}  
  \begin{tabular}{cc|cccc}
&$\tau_i$ &$H_0(\Lambda=250)$&$H_2(\Lambda=250)$&$H_0(\Lambda=600)$&$H_2(\Lambda=600)$\\ \hline
    LO&  &4.35&&0.29&\\
    \hline
\multirow{3}{*}{NLO \& N$^2$LO}&$\tau_1$&-0.90&1.99&0.57&37.87\\  
&$\tau_2$&-1.23&2.07&-1.14&7.24\\  
&$\tau_3$&-1.37&2.26&2.15&568.2\\  
 \end{tabular}
 \end{center}
  \caption{The three-body couplings $H_0$ and $H_2$ for the Gauss model and different values of the cutoff $\Lambda$. All quantities are given in MeV units. The values of $H_0$ are the same at NLO and N$^2$LO, whereas $H_2=0$ at NLO.}
\label{ValuesHGauss}
\end{table}

\begin{figure}[htb]
\subfigure{\includegraphics[width=0.49\linewidth]{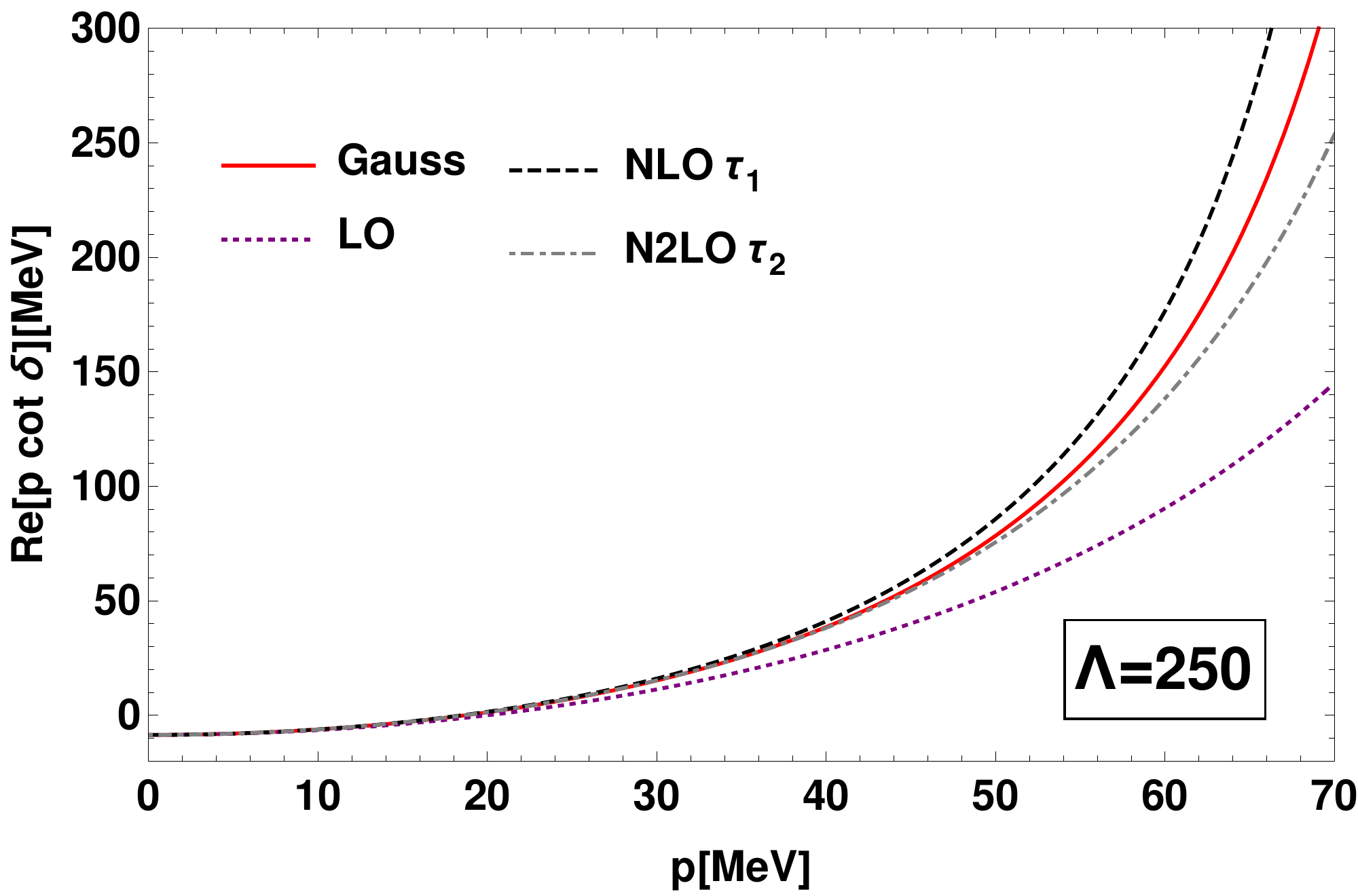}}
\hfill
\subfigure{\includegraphics[width=0.49\linewidth]{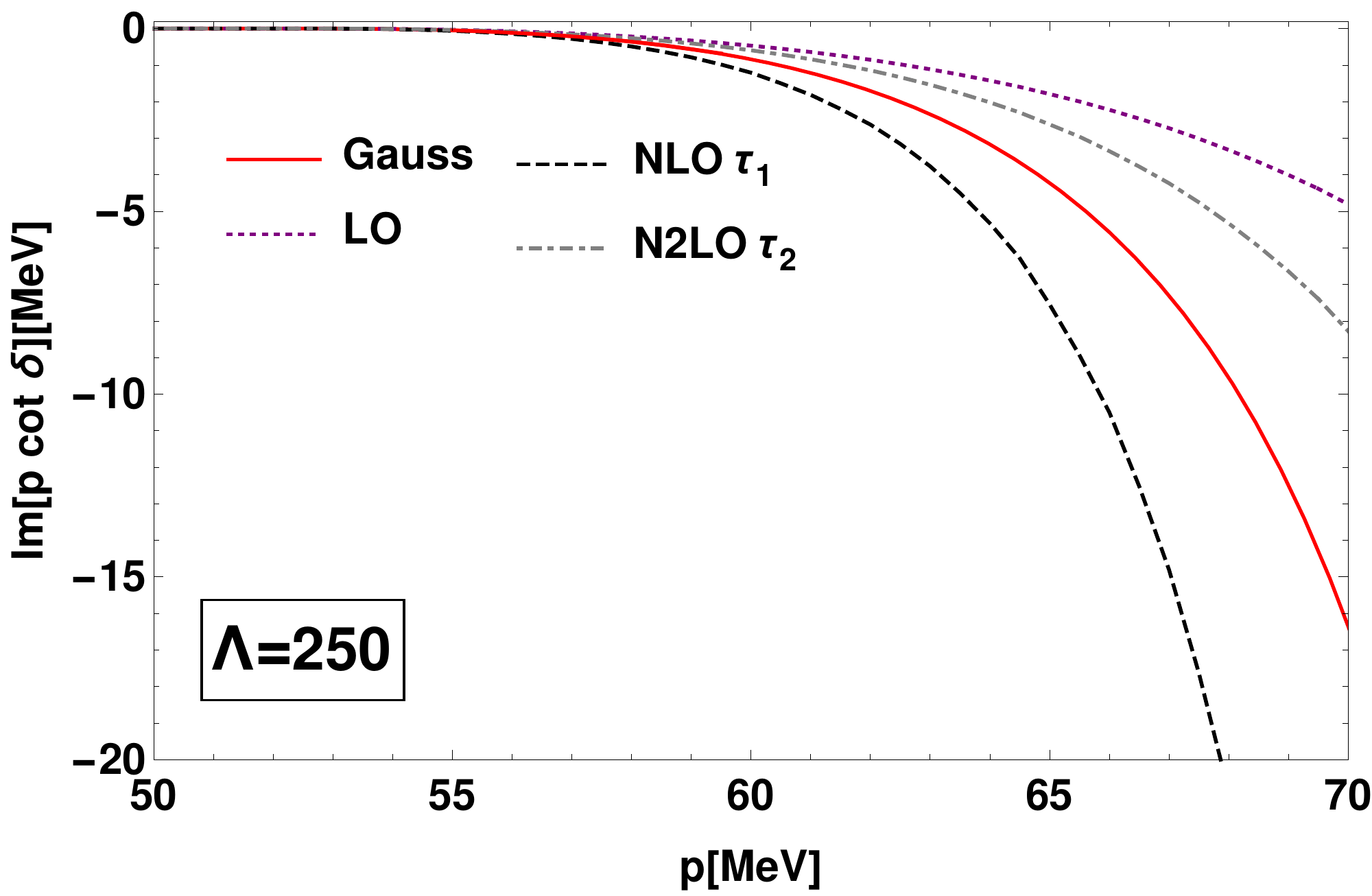}}
\caption{Numerical results for the real (imaginary) part of the quantity $p\cot\delta(p)$ in the Gauss model. Red line:
  the result obtained in the model with Gauss potential;
  in purple dotted: the LO result; in black dashed: the NLO result for $\tau_1$; in gray dot-dashed: the N$^2$LO result for $\tau_2$. The cut-off was set to the value $\Lambda=250~\text{MeV}$.}
\label{Gausskcotd}
\end{figure}

In Fig. \ref{Gausskcotd} numerical results for the Gauss model and the EFT at different orders are shown. It can be seen, that NLO and N$^2$LO are clearly better in describing the Gauss model than LO, with N$^2$LO being also better than NLO. In the right part the imaginary part is shown, the EFT results also improve order by order. It is useful to note, that since for the Yamaguchi model in the last chapter and the Gauss model here the parameters $\lambda_Y$, $\beta$ and $\lambda_G$, $E_d$ are fine-tuned to give the same $a$ and $r$, the models both exhibit a pole (zero in $\delta$) around $80 \text{ MeV}$. This results in the same problems for the EFT describing the Gauss as before. The ``window of opportunity'' is chosen between $42$ MeV and 55 MeV.

\begin{figure}[htb]
\subfigure{\includegraphics[width=0.49\linewidth]{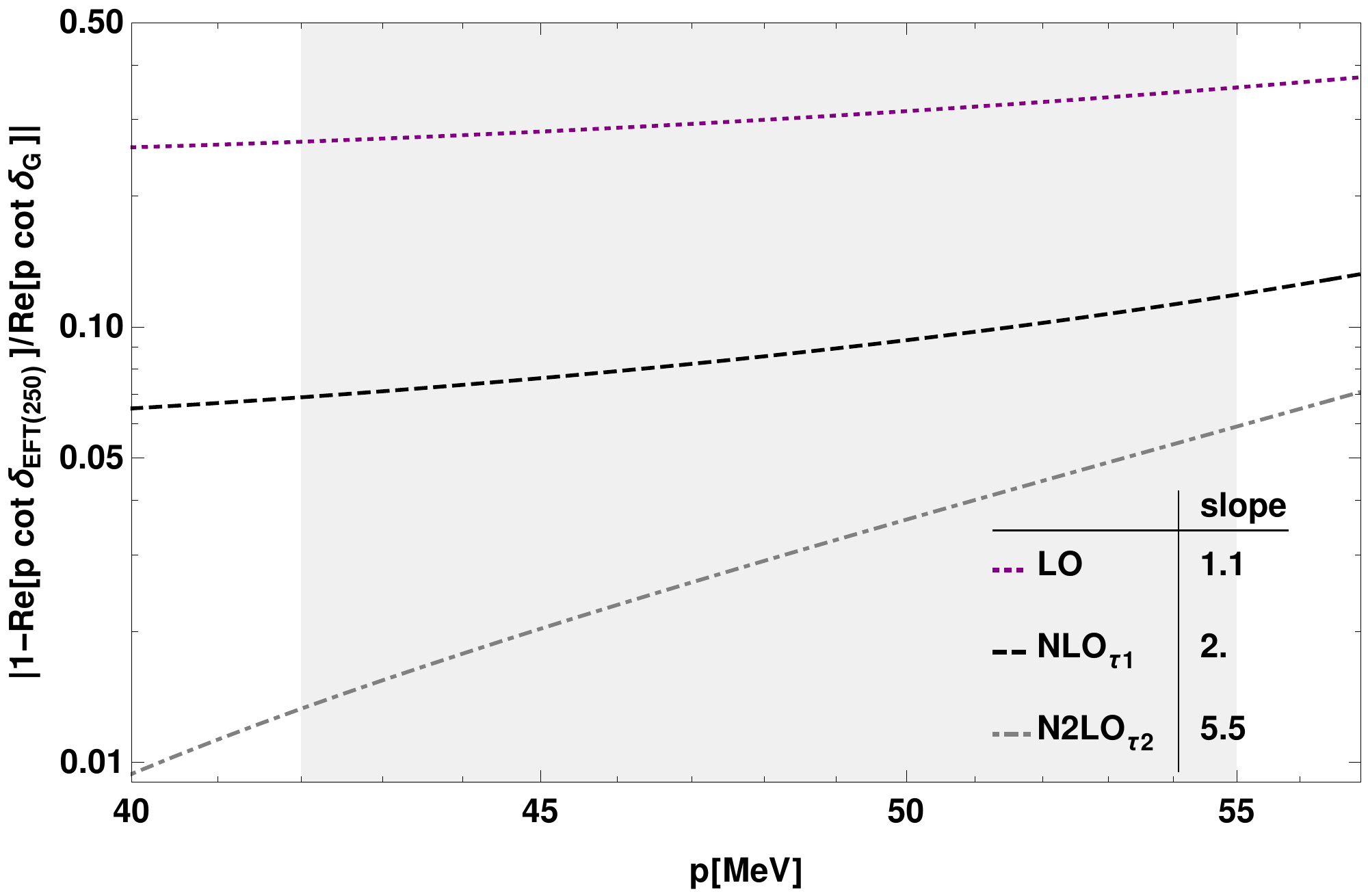}}
\hfill
\subfigure{\includegraphics[width=0.49\linewidth]{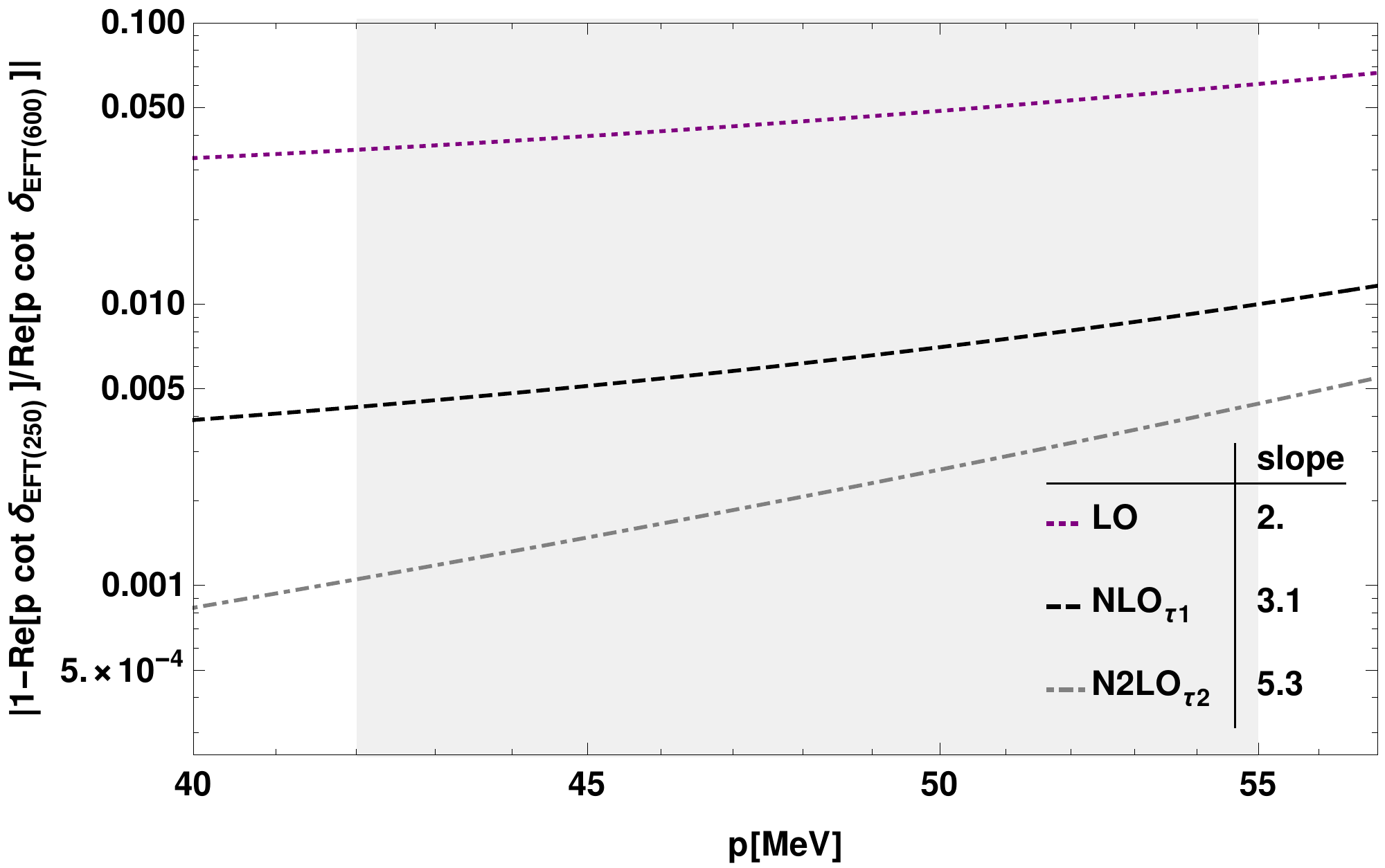}}
\caption{Lepage plot (left) for the Gauss model and consistency assessment (right) for the quantity $p\cot\delta$. The ``window of opportunity'' (shaded in gray) is chosen to be between 42 MeV and 55 MeV for all orders. }
\label{GausskcotdLepage}
\end{figure}

In the Lepage plot in Fig. \ref{GausskcotdLepage} (left) the different orders of the EFT separate nicely. The obtained slopes, shown in Table \ref{SlopeGaussLepage}, are increasing as expected. Note that the values of the slope are close to the values obtained for the Yamaguchi model. However the N$^2$LO results differ, not only is the increase of slope the slope form NLO to N$^2$LO larger than expected,\footnote{Expected is an increase of one, here the increase is around three.} also the values are larger than in the Yamaguchi case. This can be explained by the accidental zero at $p=33\text{ MeV}$. Similar to the results for the Yamaguchi model shown in the consistency assessment in Fig. \ref{YamaguchiDeltaLepage} (right) for N$^2$LO using $\tau_2(k)$ the sign of the difference is changing. In the consistency assessment in Fig. \ref{GausskcotdLepage} (right) the slopes behave as expected, the slope increases per order by approximately one.

\begin{table}[htb]
    \begin{minipage}{.5\linewidth}
      \centering
      \begin{tabular}{l|lll}
slope fit & LO & NLO & N$^2$LO\\  \hline
no sub.&1.1&&\\
 $\tau_1$ & &  2.0 & 5.6*\\
 $\tau_2$&  &  2.3 & 5.5*\\
 $\tau_3$&  &  2.5 & 5.3*\\
  \end{tabular}
    \end{minipage}%
    \begin{minipage}{.5\linewidth}
      \centering
       \begin{tabular}{l|lll}
slope fit & LO & NLO & N$^2$LO\\  \hline
no sub. &2.0&&\\
 $\tau_1$ & &  3.1 & 4.4\\
 $\tau_2$&  &  3.3 & 5.3\\
 $\tau_3$&  &  3.2 & 3.8\\
  \end{tabular}
    \end{minipage} 
    \caption{Results of the slopes for the real part of the quantity $k\cot \delta$ for the Gauss model fitted in the ``window of opportunity''. Left for the Lepage plot right for the consistency assessment. All results for the Lepage plot for N$^2$LO, marked by an asterisk,
      exhibit a accidental zero and are therefore unexpected large, compare to the Fig.~\ref{GausskcotdLepage}.}
    \label{SlopeGaussLepage}
\end{table}

To conclude, the results using the presented method to deal with the unphysical pole $k_2$ can also be used to describe the Gauss model. The description is improved order by order. The obtained slopes increase as expected as well in the Lepage plot as in the consistency assessment, where the discussed deviations are not caused by the method.

\section{Summary and Conclusions}
\label{sec:concl}

In this paper, a novel procedure for removing the contribution
from spurious poles in the three-body Faddeev equation
for pionless EFT has been proposed.
These poles emerge 
in the two particle scattering amplitudes, which enter the three-body
integral equation. Albeit the spurious
poles appear below threshold, at energies where
the EFT treatment is no more applicable, they still influence the
low-energy behavior of the particle-dimer (three-particle) amplitudes.
In the three-body integral equation the two-particle amplitudes are
evaluated at large negative energies because an
integration over all momenta is carried out. Furthermore, the
residue of these poles can have either sign, leading to the problems
with the three-particle unitarity at low energies.

In the literature, there exist different methods for treating spurious poles.
The most popular one is based on a strictly perturbative expansion
of the two-body amplitude in the range parameter(s)~\cite{Hammer:2001gh,Bedaque:1998km,Ji:2011qg,Ji:2012nj,Vanasse:2013sda}.
It will be, however, difficult to use this approach in a
finite volume for the extraction of the three-body observables
from lattice data. The reason for this is that the expansion diverges
in the vicinity of the two-particle energy levels in a finite box,
leading to more and more singular expressions at higher orders. 
  
\begin{itemize}

\item[i)]
  In the present paper, we propose a method which enables one to
  circumvent this problem, expanding only the part of the two-body
  amplitude that contains spurious poles. Such an expansion can be
  systematically carried out. Furthermore,
  in perturbation theory, the counting rules in the
  underlying EFT are closely linked to the above-mentioned expansion
  -- at a given order in the EFT counting, only first few terms in this
  expansion should be retained (the number is determined by the order
  in the EFT expansion). Adding more terms in the expansion does not lead to an increased accuracy.
  However, due to the non-perturbative character
  of the three-body integral equation, the above counting can be regarded
  merely as a rule of thumb, and the optimal number of subtractions
  should be determined in actual calculations.

\item[ii)]
  The proposal has been tested in numerical calculations in a toy
  model, using Yamaguchi and Gauss potentials in the two-body sectors.
  The results of the exact calculations have been confronted with the
  results, obtained within EFT, matched to the model parameters
  in the two- and three-body sectors. Moreover, the consistency
  assessment has been carried out, comparing the EFT results in different
  orders. In a result of these studies, a clear pattern emerges. The
  agreement with the exact calculations
 systematically improves at
  higher orders. Already at N$^2$LO, the
  exact results are reproduced very well. Moreover, expanding the
spurious pole part in the  two-body amplitude, 
  it is seen that, after few steps, the accuracy does
  not further increase when more terms are subtracted.
  This is fully in line with our expectations. The optimal
  number of the subtraction terms is slightly lower than the expectation
  from perturbation theory. This is not entirely surprising, bearing in
  mind the non-perturbative character of the three-body problem at hand.
    
\item[iii)]
  It would be extremely interesting to reformulate the three-body
  quantization condition in a finite volume as given, e.g.,
  in~\cite{Hammer:2017uqm,Hammer:2017kms,Hansen:2014eka,Hansen:2015zga,Mai:2017bge} along
  similar lines. We leave this application for a future publication.

 \end{itemize} 

\begin{acknowledgments}
  The authors would like to thank Evgeny Epelbaum and Jambul Gegelia for interesting discussions.
 M.E. was supported by a PhD fellowship from Helmholtz Forschungsakademie Hessen f\"ur FAIR (HFHF).
  M.E. and H.-W.H. were supported by Deutsche Forschungsgemeinschaft
  (DFG, German Research Foundation) --
  Project ID 279384907 -- SFB 1245.
A.R.
was  supported in part by the Deutsche Forschungsgemeinschaft
(DFG, German Research Foundation) -- Project-ID 196253076 -- TRR 110,
Volkswagenstiftung 
(grant no. 93562) and the Chinese Academy of Sciences (CAS) President's
International Fellowship Initiative (PIFI) (grant no. 2021VMB0007).
\end{acknowledgments}

\appendix
\section{Angular integral for Gauss model}
In this appendix we show how the angular integral in
the one-particle exchange diagram is calculated in the Gauss model.
The quantity $Z_G(p,q,E)$ is given by
\begin{align}
\begin{split}
  Z_G(p,q,E)&=\frac{1}{2}\int_{-1}^1 d\cos\theta_{p,q}\frac
  {\exp \left[\left({\bf p}/2+{\bf q}\right)^2/\beta^2\right]
    \exp \left[\left({\bf q}/2+{\bf p}\right)^2/\beta^2\right]}
  {E-{\bf p}^2/(2m)-{\bf q}^2/(2m) -({\bf p}+{\bf q})^2/(2m)},
\end{split}
\end{align}
this has a pole for
\begin{align}
  \left({\bf q}+\frac{\,{\bf p}}{2}\right)^2=\frac{3}{4}{\bf p}^2-mE\, ,
  \quad\mbox{or}\quad\left(\frac{\,{\bf q}}{2}+{\bf p}\right)^2=\frac{3}{4}{\bf q}^2-mE .
\end{align}
To avoid numeric problems with this pole, we add and subtract the contribution at the pole position. The kernel and the subtracted term  becomes regular at the pole. The exponential part of the added term is not angular dependent and the integral can be calculated easily: 
\begin{align}
\begin{split}
  Z_G(p,q,E)&=\frac{m}{2}\int_{-1}^1 du \frac{1}{mE-p^2-q^2-pq\, u}\\
&\times
  \Biggl(\exp\left[-\frac{q^2+p^2/4+pq\, u}{\beta^2}\right]\exp\left[-\frac{q^2/4+p^2+pq\, u}{\beta^2}\right]\\
  &-\exp\left[-\frac{3p^2/4-mE}{\beta^2}\right]\exp\left[-\frac{3q^2/4-mE}{\beta^2}\right]\Biggr)\\
  &-\frac{m}{2}\frac{1}{pq}\exp\left[-\frac{3p^2/4-mE}{\beta^2}\right]\exp\left[-\frac{3q^2/4-mE}{\beta^2}\right]\ln\left[\frac{mE-p^2-q^2-pq}{mE-p^2-q^2+pq}\right].
  \end{split}
\end{align}
This integral is well behaved at the pole position and can be calculated by standard Gaussian method.

\end{document}